\documentclass[onecolumn,nofootinbib]{revtex4-2}
\usepackage{amsfonts}
\usepackage{amsmath}
\usepackage{mathrsfs}  
\usepackage{graphicx}
\usepackage{amssymb}
\usepackage{xcolor}
\usepackage{float}
\usepackage{siunitx}
\usepackage{tensor}
\usepackage[toc]{appendix}

\newcommand{\ok}{\mbox{$k_{\perp}$}}
\newcommand{\pk}{\mbox{$k_{\parallel}$}}
\newcommand{\okk}{\mbox{$k^2_{\perp}$}}
\newcommand{\pkk}{\mbox{$k^2_{\parallel}$}}

\begin{document}

\title{On the Interaction Between Electromagnetic, 
Gravitational, and Plasma Related Perturbations 
 on LRS Class II Spacetimes}
\author{Philip Semr\'en$^{1}$ } 
\address{$^{1}$\textit{Department of Physics, Ume{\aa } University, Ume\aa, Sweden,} \\
philip.semren@umu.se}
\date{\today }

\begin{abstract}
We investigate electromagnetic, gravitational, and plasma related perturbations to first order on homogeneous and hypersurface orthogonal locally rotationally symmetric (LRS) class II spacetimes. Due to the anisotropic nature of the studied backgrounds, we are able to include a non-zero magnetic field to zeroth order.  As a result of this inclusion, we find interesting interactions between the electromagnetic and gravitational variables already to first order in the perturbations.  The equations governing these perturbations are found by using the Ricci identities, the Bianchi identities, Einstein's field equations, Maxwell's equations, particle conservation, and a form of energy-momentum conservation for the plasma components. Using a $1+1+2$ covariant split of spacetime, the studied quantities and equations are decomposed with respect to the preferred directions on the background spacetimes. After linearizing the decomposed equations around a LRS background, performing a harmonic decomposition, and imposing the cold magnetohydrodynamic (MHD) limit with a finite electrical resistivity, the system is then reduced to a set of ordinary differential equations in time and some constraints.  On solving for some of the harmonic coefficients in terms of the others, the system is found to decouple into two closed and independent subsectors. Through numerical calculations, we then observe some mechanisms for generating magnetic field perturbations, showing some traits similar to previous works using FLRW backgrounds. Furthermore, beat-like patterns are observed in the short wave length limit due to interference between gravitational waves and plasmonic modes.
\end{abstract} 

\allowdisplaybreaks

\maketitle

\section{Introduction}

Using the standard model of cosmology, based on the homogeneous and isotropic FLRW spacetimes, most astronomical observations on a cosmological scale can be explained. However, some unsolved mysteries remain. One of these mysteries pertain to the origin of the large scale magnetic fields that can be observed on the level of galaxies and clusters of galaxies. Although several mechanisms have been proposed for generating and amplifying these magnetic fields, the topic has seemingly been rather controversial \cite{Kulsrud_2008}. Amidst the controversy, a common explanation appears to be that the magnetic fields were originally created very early in the history of the universe. It is then suggested that these primordial seed fields have been amplified later on through galactic dynamo mechanisms, creating the fields we observe today \cite{Kulsrud_2008}. Although the galactic dynamo could greatly amplify the seeds, the proposed mechanisms that generate the seeds may be too weak, or, in some cases, lie uncomfortably close to the minimum strength requirement. In addition to the strength requirement, there are also stringent requirements on the coherence length of the seeds, further limiting the possible candidates \footnote{For a more detailed account of these matters, the reader is referred to \cite{Tsagas_2003}}. 

To increase the number of possible candidates, one way around the stringent requirements would be if some mechanism existed for amplifying the seeds after they have been created, but acting before the galactic dynamo starts. Some possible mechanisms of this kind have already been proposed. Examples include mechanisms based on gravitational wave interactions \cite{Tsagas_2003, Betschart_2005, Zunckel_2006}, and velocity perturbations in the cosmological plasma \cite{Zunckel_2006}. Although some of the mechanisms for amplifying the seed fields have been suggested to be too weak \cite{Fenu_2009, Kuroyanagi_2010, Mongwane_2012}, the prospect of a mechanism based on the dynamics of a self-gravitating plasma and classical general relativity is still enticing, since it would not explicitly depend on any additional exotic physics. Therefore, it is still of great interest to study the interactions between a cosmological plasma and general relativistic effects. 

The studies of self-gravitating plasmas in a cosmological scenario seem to have mainly used a perturbative approach with the FLRW spacetimes as zeroth order backgrounds \cite{Tsagas_2003, Betschart_2005,Zunckel_2006, Fenu_2009, Kuroyanagi_2010, Mongwane_2012, BetschartDunsby_2004}. However, due to the isotropy of the FLRW spacetimes, it is essentially impossible to have a non-zero magnetic field 3-vector on the background, since that field would otherwise define a preferred spatial direction. To circumvent this problem, several different approaches have been used. One approach, which is usually referred to as the weak-field approximation, is to assume that the background magnetic field is sufficiently small so that it does not contribute to the energy-momentum tensor, which involves squares of the magnetic field \cite{Tsagas_2003}. With this assumption, the magnetic field should not affect the underlying geometry, and hence, in a sense, not violate the isotropy of the background spacetime. However, this approach runs in to some formal problems, since the generated magnetic field perturbations that are explicitly studied are not gauge invariant in the strict mathematical sense of the Stewart-Walker lemma \cite{Betschart_2005, Stewart_1974}. 

To better avoid interference from gauge related degrees of freedom, second order schemes have also been used. The magnetic field is then introduced as a first order perturbation rather than on the background \cite{Betschart_2005, Zunckel_2006, Mongwane_2012}. The interaction between the gravitational effects and the electromagnetic fields then becomes a second order quantity, since this involves products of the electromagnetic fields and other first order variables. However, some care is still needed, as it is not strictly correct to integrate the second order results for the gauge invariant variables to obtain explicit expressions for the generated magnetic field, since this quantity is not gauge invariant \cite{Mongwane_2012}. Hence, one is restricted to interpreting various gauge invariant quantities to draw any conclusion about the generated magnetic fields \cite{Mongwane_2012}. 

Finally, a more drastic approach, which is the one that will be used here, is to use an anisotropic background model instead of the FLRW spacetimes. Then there is no longer any isotropy problem, and a non-zero magnetic field to zeroth order is allowed without causing severe problems with the background geometry.  In this paper, the aim is therefore to investigate if it is possible to get interesting gauge invariant interactions already to first order in the perturbations if we use a certain class of anisotropic backgrounds instead of the FLRW spacetimes. More specifically, we will consider backgrounds that are homogeneous, hypersurface orthogonal, and belong to the locally rotationally symmetric (LRS) class II of spacetimes, which allow for a preferred spatial direction.  

First order perturbations on LRS class II spacetimes have been studied previously, both in the perfect fluid case  \cite{Bradley_2011, Keresztes_2015, Bradley_2017, Tornkvist_2019}, and when including dissipative effects  \cite{Bradley_2021}. However, electromagnetic perturbations on general LRS class II spacetimes have, to our knowledge, mainly been considered as test fields, without any electromagnetic fields on the background \cite{BetschartClarksson_2004, Burston_2008}. Hence, it is still interesting to investigate electromagnetic perturbations on LRS class II spacetimes when including more interactions with gravitational and matter related perturbations.  

Our approach to study these interactions will follow the same steps as in \cite{Bradley_2021}, but instead of specializing the equations to describe one-component dissipative fluids, we will consider self-gravitating plasmas. Thus, we will make use of the results for general energy-momentum tensors presented in \cite{Bradley_2021}, but we will then close the general system by imposing Maxwell's equations, particle conservation, and equations describing energy-momentum conservation for the plasma components. In doing so, this article reproduces the methods and equations from \cite{Semren_2021} and an upcoming contribution to the proceedings from the sixteenth Marcel Grossmann meeting on general relativity \cite{Semren_2022}, but extends these accounts with additional numerical results and discussions. 

\section{Spacetime dynamics} \label{sec:Spacetime dynamics}

To determine the behavior of both the zeroth order background and the first order perturbations, we will follow \cite{Clarkson_2007} and make use of the Ricci identities, the Bianchi identities, and Einstein's field equations. In doing so, we will assume that the studied spacetimes have two preferred directions as specified by the vector fields $\tensor{u}{^a} $ and $\tensor{n}{^a} $. The field $\tensor{u}{^a} $ is assumed to be timelike and normalized as  $\tensor{u}{^a} \tensor{u}{_a} = -1 $, whilst  $\tensor{n}{^a} $ is spacelike with $\tensor{n}{^a}\tensor{n}{_a} = 1$ \footnote{In the following we will use units so that $c = 8\pi G/c^4 = 1$. When describing electromagnetic quantities, we will use Lorentz-Heaviside units.}. As for the physical interpretation of these directions, $\tensor{u}{^a} $ denotes the 4-velocity of some fundamental observer, whilst $\tensor{n}{^a}$ defines the direction with respect to which the backgrounds are assumed to be locally rotationally symmetric.  It is then natural to make use of the Ricci identities for these preferred vector fields,
\begin{align}
 2\tensor{\nabla}{_{[a}}\tensor{\nabla}{_{b]}}\tensor{u}{_c} - \tensor{R}{_a_b_c_d}\tensor{u}{^d} &= 0,  \label{ps:eq1}\\
 2\tensor{\nabla}{_{[a}}\tensor{\nabla}{_{b]}}\tensor{n}{_c} - \tensor{R}{_a_b_c_d}\tensor{n}{^d} &= 0.
\end{align}
Together with the contracted Bianchi identities 
\begin{align}
 2\tensor{\nabla}{_{[e}}\tensor{R}{_{d]}_b} + \tensor{\nabla}{_a}\tensor{R}{^a_b_d_e} &=0, \\
\tensor{\nabla}{^a}\tensor{G}{_a_b} &= 0, 
\end{align}
and Einstein's field equations
\begin{equation}
	\tensor{G}{_a_b} = \tensor{T}{_a_b} - \Lambda\tensor{g}{_a_b}, 
\end{equation}
these relations provide most of the necessary equations. Here $\tensor{G}{_a_b}\equiv \tensor{R}{_a_b} - \tensor{g}{_a_b}R/2$ is the usual Einstein tensor, whilst $\tensor{T}{_a_b}$ is the energy-momentum tensor, which could be general at this point. However, with a general energy-momentum tensor, these equations will not be enough to get a closed system \cite{Bradley_2021}. Therefore, to arrive at a closed system, we will here specialize the energy-momentum tensor to describe a plasma together with some electromagnetic fields. Hence we assume that 
\begin{equation}
\tensor{T}{^a^b} =\tensor{{T_{(F)}}}{^a^b} + \tensor{{T_{(EM)}}}{^a^b}, 
\end{equation}
where $\tensor{{T_{(F)}}}{^a^b}$ is due to the plasma whilst $\tensor{{T_{(EM)}}}{^a^b}$ comes from the macroscopic electromagnetic fields. The electromagnetic part can in turn be written as 
\begin{equation}
\tensor{{T_{(EM)}}}{^a^b}  = \tensor{F}{^a_c}\tensor{F}{^b^c} - \frac{1}{4}\tensor{g}{^a^b}\tensor{F}{^c^d}\tensor{F}{_c_d},
\end{equation}
in terms of the Faraday tensor $\tensor{F}{_a_b}$, which satisfies the usual Maxwell's equations 
\begin{align}
\tensor{\nabla}{_{[a}}\tensor{F}{_b_{c]}} &= 0, \\ 
\tensor{\nabla}{_b}\tensor{F}{^a^b} &= \tensor{j}{^a},
\end{align}
where $\tensor{j}{^a}$ is the 4-current density. 

As for the plasma contribution, we will follow the formalism in \cite{Marklund_2003}  and assume that this contribution consists of separate contributions from each plasma component, so that
\begin{equation}
\tensor{{T_{(F)}}}{^a^b} = \sum_{i}\tensor{{T_{(i)}}}{^a^b},
\end{equation}
where the $i$th component satisfies the following energy-momentum conservation equation
\begin{align}
\tensor{\nabla}{_b}\tensor{{T_{(i)}}}{^a^b} &= \tensor{F}{^a^b}\tensor{ {j_{(i)}} }{_b} + \tensor{{I_{(i)}}}{^a}, \label{ps:eq11}
\end{align}
with 
\begin{equation} 
\sum_i  \tensor{{I_{(i)}}}{^a} = 0. \label{ps:eq12}
\end{equation}
Here $\tensor{{I_{(i)}}}{^a}$ includes interactions such as collisions between the different fluid components, whilst $\tensor{ {j_{(i)}} }{^a}$ is the 4-current density of the $i$th component. The current density satisfies the relations
\begin{align}
\tensor{{j_{(i)}}}{^a} &= \rho_{(i)}\tensor{{u_{(i)}}}{^a}, \label{ps:eq13}\\ 
\tensor{j}{^a} &= \sum_{i}\tensor{ {{j_{(i)}}} }{^a},
\end{align}
where $\rho_{(i)} =  q_{c(i)}{\mathcal{N}_{(i)}}$ is the charge density, $q_{c(i)}$ is the charge, ${\mathcal{N}_{(i)}}$ is the number density, and $\tensor{ {u_{(i)}} }{^a}$ is the the 4-velocity of the $i$th component. In the following, we will also write the interaction terms and the total current density as
\begin{align}
\tensor{{I_{(i)}}}{^a} &=  \varepsilon_{(i)}\tensor{u}{^a} + \tensor{{f_{(i)}}}{^a}, \label{ps:eq15}  \\
~  
\tensor{j}{^a} &= \rho\tensor{u}{^a} + \tensor{J}{^a}, \label{ps:eq16} 
\end{align}
where $\tensor{J}{^a}$ and $\tensor{{f_{(i)}}}{^a}$ are orthogonal to $\tensor{u}{^a}$. Furthermore, we will assume that each fluid component can be described as a perfect fluid, so that 
\begin{equation}
 \tensor{{T_{(i)}}}{^a^b} = \left(\mu_{(i)} + p_{(i)}\right)\tensor{{u_{(i)}}}{^a}\tensor{{u_{(i)}}}{^b} + p_{(i)}\tensor{g}{^a^b}, 
 \label{ps:eq17}
\end{equation}
where $\mu_{(i)}$ and $p_{(i)}$ are the energy density and pressure relative to the component rest frame, as defined by the 4-velocity $\tensor{{u_{(i)}}}{^a}$. This 4-velocity can in turn be written as 
\begin{equation}
\tensor{ {u_{(i)}} }{^a} = \gamma_{(i)}\left(\tensor{u}{^a} + \tensor{ {v_{(i)}} }{^a} \right),
\end{equation}
where
\begin{equation}
\tensor{u}{^a}\tensor{{v_{(i)}}}{_a} = 0, \quad \gamma_{(i)}= \frac{1}{\sqrt{1-{v_{(i)}}^2}}, \quad {v_{(i)}}^2 \equiv \tensor{ {v_{(i)}} }{_a}\tensor{ {v_{(i)}} }{^a}.
\end{equation}
We then also impose particle conservation for each fluid component by requiring that 
\begin{align}
\tensor{\nabla}{_a}\left( {\mathcal{N}_{(i)}} \tensor{ {u_{(i)}} }{^a}\right) &= 0.
\end{align}

Finally, after specifying a microscopic description of $\tensor{{I_{(i)}}}{^a}$, equations of state $p_{(i)} = p_{(i)}(\mu_{(i)}, \mathcal{N}_{(i)})$, and choosing a specific frame, we now have enough equations to get a closed system. To get a simple, yet still informative, system, we will later on assume that the plasma is sufficiently cold so that $p_{(i)} = 0$. Then we will also assume that the interactions between the fluid components can be described with a single scalar electrical resistivity, $\eta$, in the magnetohydrodynamic (MHD) approximation. In this approximation, the plasma is described in terms of some collective variables instead of the individual component variables. Therefore, after linearizing and harmonically decomposing the equations, we will write them in terms of the harmonic coefficients of the following total fluid energy density and average 3-velocity
\begin{align}
{\mu_{(F)}} &\equiv \sum_{i}{\mu_{(i)}}, \\
{\mu_{(F)}}\tensor{ {v_{(F)}} }{^a} &\equiv \sum_{i}{\mu_{(i)}}\tensor{ {v_{(i)}} }{^a}.
\end{align}
Using Ohm's law, the total 3-current density can then be written as 
\begin{equation}
\tensor{J}{^{a}} = \frac{1}{\eta}\left(\tensor{E}{^a} + \tensor{\epsilon}{^a^b^c}\tensor{ {v_{(F)}} }{_b}\tensor{B}{_c}\right),
\label{eq:OhmsLaw}
\end{equation} 
where $\tensor{E}{^a} $ and $\tensor{B}{^a} $ are the electric and magnetic fields defined through
\begin{equation}
\tensor{F}{_a_b} = \tensor{\epsilon}{_a_b_c}\tensor{B}{^c} + 2\tensor{u}{_{[a}}\tensor{E}{_{b]}},
\end{equation}
with $\tensor{\epsilon}{_a_b_c} \equiv \tensor{\eta}{_d_a_b_c}\tensor{u}{^d} \equiv 4!\sqrt{-g}\tensor{\delta}{^0_{[d}}\tensor{\delta}{^1_a}\tensor{\delta}{^2_b}\tensor{\delta}{^3_{c]}}\tensor{u}{^d}$. However, the cold MHD assumption will only be imposed after linearizing and harmonically decomposing the equations. Until then, we will continue using the more general multifluid description with non-zero $p_{(i)}$ and with interactions and current densities determined by Eqs.~(\ref{ps:eq12})–(\ref{ps:eq16})  .

\section{Covariant Splits of Spacetime}

As the studied spacetimes are assumed to have two preferred directions, it is natural to decompose the studied quantities with respect to those directions. Since the vector fields defining these directions are physical objects, the decompositions respect the tensorial properties of the decomposed variables. Hence they are usually referred to as covariant splits of spacetime. 

\subsection{$1+3$}
Starting with a $1+3$ split with respect to the timelike vector field $\tensor{u}{^a}$, we write the metric as 
\begin{equation}
	\tensor{g}{_a_b} = \tensor{U}{_a_b}+ \tensor{h}{_a_b},
	\label{ps:eq18}
\end{equation}
where $\tensor{U}{_a_b} = -\tensor{u}{_a}\tensor{u}{_b}$ is a projection tensor along $\tensor{u}{^a}$ whilst $\tensor{h}{_a_b}$ projects onto the 3-dimensional hypersurfaces orthogonal to $\tensor{u}{^a}$ \cite{Ellis_1998}. Using Eq.~\eqref{ps:eq18}, all relevant objects can then be decomposed relative to $\tensor{u}{_a}$ and the 3-dimensional hypersurfaces. Some of the quantities from Sec.~\ref{sec:Spacetime dynamics} have already been decomposed in this manner, such as the component 4-velocities, the total current density, and the Faraday tensor.  However, it remains to fully decompose the total energy-momentum tensor and the Riemann tensor. Performing this decomposition we encounter the following projected variables
\begin{align}
\mu & \equiv 	\tensor{T}{^c^d}\tensor{u}{_c}\tensor{u}{_d}, \\
	\tensor{q}{_a} & \equiv -\tensor{T}{^c^d}\tensor{u}{_c}\tensor{h}{_d_{a}}, \\
	p &\equiv \frac{1}{3}\tensor{T}{^c^d}\tensor{h}{_c_d},  \\
	\tensor{\pi}{_a_b} &\equiv \left(\tensor{h}{_{(a}^c}\tensor{h}{_{b)}^d} - \frac{1}{3}\tensor{h}{^c^d}\tensor{h}{_a_b} \right)\tensor{T}{_c_d} \equiv \tensor{T}{_{\langle a}_{b\rangle}}, \\
\tensor{E}{_a_b} &\equiv  \tensor{C}{_a_c_b_d}\tensor{u}{^c}\tensor{u}{^d}, \\
 \tensor{H}{_a_b} &\equiv \frac{1}{2}\tensor{\epsilon}{_a^e^f}\tensor{C}{_e_f_b_d}\tensor{u}{^d},
\end{align}
where $\mu$ is the energy density, $\tensor{q}{^a}$ is the energy flux orthogonal to  $\tensor{u}{^a}$, $p$ is the isotropic pressure, $\tensor{\pi}{_a_b}$ is the trace-free anisotropic pressure, whilst $\tensor{E}{_a_b}$  and $\tensor{H}{_a_b}$ are the electric and magnetic parts of the Weyl tensor $\tensor{C}{_a_b_c_d}$ \cite{Ellis_1998}. When defining the anisotropic pressure $\tensor{\pi}{_a_b}$, we have introduced a notation with angular brackets to denote the projected, symmetric, and trace-free (PSTF) part of a tensor relative to the projection tensor $\tensor{h}{_a_b}$. 

We also obtain some new projected variables when considering the covariant derivative of the preferred vector field. Defining the time derivative
\begin{equation}
	\tensor{\dot{T}}{^a^b^{\ldots}_c_d_{\ldots}} \equiv \tensor{u}{^e}\tensor{\nabla}{_e}\tensor{T}{^a^b^{\ldots}_c_d_{\ldots}},
\end{equation}
and the spatial derivative fully projected onto the 3-dimensional hypersurfaces
\begin{equation}
	\tensor{D}{_e}\tensor{T}{^a^b^{\ldots}_c_d_{\ldots}} \equiv \tensor{h}{^a_f}\tensor{h}{^b_g}\cdots\tensor{h}{_c^h}\tensor{h}{_d^i}\cdots\tensor{h}{_e^j}\tensor{\nabla}{_j}\tensor{T}{^f^g^{\ldots}_h_i_{\ldots}},
\end{equation} 
we can write the covariant derivative of $\tensor{u}{^a}$ as 
 \begin{equation}
	\tensor{\nabla}{_a}\tensor{u}{_b} = -\tensor{u}{_a}\tensor{\dot{u}}{_b} + \frac{1}{3}\Theta\tensor{h}{_a_b} + \tensor{\omega}{_a_b} + \tensor{\sigma}{_a_b}.
\end{equation}
Here $\Theta \equiv  \tensor{D}{^a}\tensor{u}{_a}$ is the expansion rate, $\tensor{\omega}{_a_b} \equiv \tensor{D}{_{[a}}\tensor{u}{_{b]}}$ is the vorticity, and $\tensor{\sigma}{_a_b} \equiv \tensor{D}{_{\langle a}}\tensor{u}{_{b\rangle}}$ is the rate of shear \cite{Ellis_1998}. Instead of working directly with the vorticity tensor, we will write it in terms of the vorticity vector
\begin{equation}
	\tensor{\omega}{^a} \equiv \frac{1}{2}\tensor{\epsilon}{^a^b^c}\tensor{\omega}{_b_c}.
\end{equation}

\subsection{$1+1+2$}

After having decomposed all relevant variables with respect to $\tensor{u}{^a}$, we can go one step further and also decompose the 3-dimensional hypersurfaces with respect to $\tensor{n}{^a}$, leading to a so-called $1+1+2$ covariant split \cite{Clarkson_2007}. For this purpose, we write the projection tensor $\tensor{h}{_a_b}$ as 
\begin{equation}
	\tensor{h}{_a_b} = \tensor{N}{_a_b} + \tensor{n}{_a}\tensor{n}{_b}.
\label{ps:eqhdecomp}
\end{equation}  
where $\tensor{N}{_a_b}$ projects onto the 2-dimensional sheets orthogonal to both $\tensor{u}{^a}$ and $\tensor{n}{^a}$.  Using Eq.~\eqref{ps:eqhdecomp}, the relevant 3-vectors can then be decomposed as
\begin{alignat}{2}
	\tensor{q}{^a} &=  Q\tensor{n}{^a} + \tensor{Q}{^a}, \quad & \tensor{B}{^a} &= \mathfrak{B}\tensor{n}{^a}+\tensor{\mathfrak{B}}{^a}, \\
	~
	\tensor{\omega}{^a} &= \Omega\tensor{n}{^a} + \tensor{\Omega}{^a}, \quad & \tensor{E}{^a} &= \mathfrak{E}\tensor{n}{^a}+\tensor{\mathfrak{E}}{^a}, \\
	~ 
	\tensor{\dot{u}}{^a} &= \mathcal{A}\tensor{n}{^a} + \tensor{\mathcal{A}}{^a}, \quad & \tensor{J}{^a} &= \mathcal{J}\tensor{n}{^a}+\tensor{\mathcal{J}}{^a}, \\
	~ 
	\tensor{\dot{n}}{^a} &= \mathcal{A}\tensor{u}{^a} + \tensor{\alpha}{^a}, \quad &\tensor{ {f_{(i)}} }{^{a}} &= {\mathcal{F}_{(i)}}\tensor{n}{^a}+\tensor{{\mathcal{F}_{(i)}}}{^a}, \\
	\tensor{ {v_{(i)}} }{^{a}} &= {\mathcal{V}_{(i)}}\tensor{n}{^a}+\tensor{{\mathcal{V}_{(i)}}}{^a}, \quad & \tensor{ {v_{(F)}} }{^{a}} &= {\mathcal{V}_{(F)}}\tensor{n}{^a}+\tensor{{\mathcal{V}_{(F)}}}{^a} ,
\end{alignat}
Occasionally, we will use a notation with a bar over an index to denote a projection with $N\indices{_a_b}$, so that $\psi\indices{_{\bar{a}}} =N\indices{_a^b}\psi_b$.

As for PSTF 3-tensors $\tensor{\psi}{_a_b}$, these are decomposed as 
\begin{equation}
	\tensor{\psi}{_a_b} = \Psi\left(\tensor{n}{_a}\tensor{n}{_b} - \frac{1}{2}\tensor{N}{_a_b} \right) + 2\tensor{n}{_{(a}}\tensor{\Psi}{_{b)}} + \tensor{\Psi}{_{a}_{b}},
\end{equation}
where $\tensor{n}{^a}\tensor{\Psi}{_a} = 0$ and 
\begin{equation}
\tensor{\Psi}{_{a}_{b}} =\left(\tensor{N}{_{(a}^c}\tensor{N}{_{b)}^d}  - \frac{1}{2}\tensor{N}{_{a}_b}\tensor{N}{^c^d}\right) \tensor{\psi}{_c_d} \equiv \tensor{\psi}{_{\{a}_{b\}}}. 
\end{equation}
Here we have used a notation with curly brackets to denote the PSTF-part of a tensor with respect to the projection tensor $\tensor{N}{_a_b}$. For the PSTF 3-tensors relevant here, we therefore write
\begin{align}
\tensor{\pi}{_a_b} =~&\Pi\left(\tensor{n}{_a}\tensor{n}{_b} - \frac{1}{2}\tensor{N}{_a_b} \right) + 2\tensor{n}{_{(a}}\tensor{\Pi}{_{b)}} + \tensor{\Pi}{_{a}_{b}},\\
	\tensor{E}{_a_b} =~& \mathcal{E}\left(\tensor{n}{_a}\tensor{n}{_b} - \frac{1}{2}\tensor{N}{_a_b} \right) + 2\tensor{n}{_{(a}}\tensor{\mathcal{E}}{_{b)}} + \tensor{\mathcal{E}}{_{a}_{b}},\\
	~
	\tensor{H}{_a_b} =~& \mathcal{H}\left(\tensor{n}{_a}\tensor{n}{_b} - \frac{1}{2}\tensor{N}{_a_b} \right) + 2\tensor{n}{_{(a}}\tensor{\mathcal{H}}{_{b)}} + \tensor{\mathcal{H}}{_{a}_{b}},\\
	~
	\tensor{\sigma}{_a_b} =~& \Sigma\left(\tensor{n}{_a}\tensor{n}{_b} - \frac{1}{2}\tensor{N}{_a_b} \right) + 2\tensor{n}{_{(a}}\tensor{\Sigma}{_{b)}} + \tensor{\Sigma}{_{a}_{b}}.
\end{align}

In a similar manner as before, some new projected variables are also obtained when considering the spatial derivative of $\tensor{n}{^a}$. Defining the spatial gradient along $\tensor{n}{^a}$,
\begin{equation}
 	\tensor{\hat{T}}{^a^b^{\ldots}_c_d_{\ldots}} \equiv \tensor{n}{^e}\tensor{D}{_e}\tensor{T}{^a^b^{\ldots}_c_d_{\ldots}},
\end{equation}
and the spatial derivative fully projected onto the 2-sheets,
\begin{equation}
	\tensor{\delta}{_e}\tensor{T}{^a^b^{\ldots}_c_d_{\ldots}} \equiv \tensor{N}{^a_f}\tensor{N}{^b_g}\cdots\tensor{N}{_c^h}\tensor{N}{_d^i}\cdots\tensor{N}{_e^j}\tensor{D}{_j}\tensor{T}{^f^g^{\ldots}_h_i_{\ldots}}, 
\end{equation}
the spatial derivative of $\tensor{n}{^a}$ can be written as
\begin{equation}
\tensor{D}{_a}\tensor{n}{_b} = \tensor{n}{_a}\tensor{a}{_b} + \frac{1}{2}\phi\tensor{N}{_a_b} +  \xi\tensor{\epsilon}{_a_b} + \tensor{\zeta}{_a_b},
\end{equation} 
where $\tensor{\epsilon}{_a_b} \equiv \tensor{\epsilon}{_a_b_c}\tensor{n}{^c}$. Here $\phi \equiv \tensor{\delta}{^c}\tensor{n}{_c}$ describes an expansion of the 2-sheets,  $\xi\tensor{\epsilon}{_a_b} \equiv  \tensor{\delta}{_{[a}}\tensor{n}{_{b]}}$ describes a twist of these sheets, $\tensor{\zeta}{_a_b} \equiv \tensor{\delta}{_{\{a}}\tensor{n}{_{b\}}}$ is the so called distortion, and $\tensor{a}{_a} \equiv  \tensor{\hat{n}}{_a}$ \cite{Clarkson_2007}. 

After having decomposed the relevant variables, we also need to decompose the equations from Sec.~\ref{sec:Spacetime dynamics}. This involves decomposing the covariant derivatives of scalars, 2-vectors, PSTF 2-tensors,  $\tensor{\epsilon}{_a_b}$ and  $\tensor{N}{_a_b}$. After inserting all of the decompositions into the equations, they are then contracted with various combinations of $\tensor{u}{^a}$,$\tensor{n}{^a}$, $\tensor{\epsilon}{_a_b}$ and  $\tensor{N}{_a_b}$ to extract the individual equations for the projections. However, since this is a rather long process, we will omit the details here and continue with the perturbative approach. 

\section{Background Spacetimes}\label{sec:BackgroundSpacetimes}

We begin by considering the equations to zeroth order. As such, we need to specify the properties of the background spacetimes to be studied. These are chosen to be homogeneous and hypersurface orthogonal members of LRS class II, which is characterized by that $\xi$, $\tensor{\omega}{_a_b}$, and $\tensor{H}{_a_b}$ are all identically zero \cite{Stewart_1968, BetschartClarksson_2004}. To be able to use the same type of harmonic decomposition for all of the studied spacetimes, we will also impose the requirement that $\phi = 0$ \cite{Bradley_2017}. Additionally, it is also assumed that the fluid velocities $\mathcal{V}_{(i)}$, and the interaction terms $\mathcal{F}_{(i)}$ and  $\varepsilon_{(i)}$ all vanish identically to zeroth order. Due to all of these assumptions, the only relevant non-zero dynamical variables to zeroth order are
\begin{equation}
S^{(0)} = \left \{\Theta, \Sigma,{\mu_{(i)}}, {\mathcal{N}_{(i)}}, \mathfrak{B}, \mathcal{E},\mu, p, \Pi, {p_{(i)}}  \right\}.
\end{equation}
Setting all other variables to zero in the equations from the previous section and using the spatial homogeneity of the background, the quantities in $S^{(0)}$ are found to satisfy the relations
\begin{align}
\dot{\Theta}&=-\frac{\Theta ^{2}}{3}-\frac{3\Sigma ^{2}}{2}-\frac{1}{2}\left(
\mu +3p\right) +\Lambda, \label{eq:Theta_dot} \\
\dot{\Sigma}&=\frac{2}{3}\left( \mu +\Lambda \right) +\frac{\Sigma ^{2}}{2} 
-\Sigma \Theta -\frac{2\Theta ^{2}}{9}+\Pi,  \label{eq:Sigma_dot}\\
{\dot{\mu}_{(i)}} &= - \left({\mu_{(i)}} + {p_{(i)}} \right) \Theta,  \\
{\dot{\mathcal{N}}_{(i)}} &= -  \mathcal{N}_{(i)}\Theta,  \\
\dot{\mathfrak{B}} &=  \left(\Sigma - \frac{2\Theta}{3}\right)\mathfrak{B}, \label{eq:b_dot}
\end{align}
where 
\begin{align}
3\mathcal{E} &=-2\left( \mu +\Lambda \right) -3\Sigma ^{2}+\frac{2\Theta
^{2}}{3}+\Sigma \Theta -\frac{3\Pi}{2}, \\
\mu &= \sum_{i} {\mu_{(i)}} + \frac{\mathfrak{B}^2 }{2}, \quad p = \sum_{i} {p_{(i)}} + \frac{\mathfrak{B}^2}{6}, \quad \Pi = -\frac{2\mathfrak{B}^2}{3}, \\
0&=\sum_{i}q_{c(i)} {\mathcal{N}_{(i)}} ,
\end{align}
It should here be noted that we need an additional equation for the pressures ${p_{(i)}}$ to get a closed system. Hence, on specifying equations of state ${p_{(i)}} =  {p_{(i)}}({\mu_{(i)}},{\mathcal{N}_{(i)}})$ and a set of initial conditions, the behavior of the background spacetime is fully determined by the system above.   

Furthermore, on introducing coordinates, the line element for this class of spacetimes can be written as  \cite{Stewart_1968, Bradley_2017}
\begin{equation}
ds^{2}=-dt^{2}+a_{1}^{2}\left( t\right) dz^{2}+a_{2}^{2}\left( t\right)
\left( d\vartheta ^{2}+f_{\kappa}(\vartheta) d\varphi ^{2}\right),
\end{equation}
where the scale factors $a_1$ and $a_2$ are related to the expansion rate and the shear through the relations 
\begin{equation}
\Theta =\frac{\dot{a}_{1}}{a_{1}}+2\frac{\dot{a}_{2}}{a_{2}}, \quad \Sigma =\frac{2}{3}\left( \frac{\dot{a}_{1}}{a_{1}}-\frac{\dot{a}_{2}}{a_{2}}
\right),  \label{eq:ThetaSigmaToa}
\end{equation}
or 
\begin{equation}
\frac{\dot{a}_{1}}{a_{1}} = \Sigma + \frac{\Theta}{3}, \quad \frac{\dot{a}_{2}}{a_{2}} = -\frac{1}{2}\left(\Sigma - \frac{2\Theta}{3}\right). \label{eq:adot}
\end{equation}
As for the precise nature of the function $f_{\kappa}(\vartheta)$, this will depend on the 2D curvature  \cite{Keresztes_2015}
\begin{equation}
\mathcal{R} = \frac{2\kappa}{a_2^2} =  2\left(\mu + \Lambda\right) + \frac{3\Sigma^2}{2} - \frac{2\Theta^2}{3}.
\end{equation}
For $\kappa = 0$ the 2-sheets are flat with either $f_{0}(\vartheta) = 1$ or $f_{0}(\vartheta)=\vartheta^2$. If instead $\kappa = 1$, the sheets are closed with $f_{1}(\vartheta) = \sin^2\vartheta$. Lastly, if $\kappa = -1$ the sheets are open with $f_{-1}(\vartheta) = \sinh^2\vartheta$.

\section{Perturbations}\label{sec:Perturbations}

With the zeroth order spacetime specified, we now consider first order perturbations on this background. To avoid unphysical modes due to the implicit mapping between the background spacetime and the perturbed manifold, we will describe the perturbations in terms of gauge invariant variables \cite{Ellis_1998}. Making use of the Stewart-Walker lemma, the gauge invariant variables to first order are chosen to be quantities that vanish on the background \cite{Stewart_1974}. Hence, quantities such as  $\tensor{\mathcal{H}}{_{a}_{b}}$ and $ \tensor{\mathfrak{B}}{_a}$, which are naturally zero on the background, are gauge invariant to first order. However, not all quantities are zero on the background. Thus, to represent the perturbations of the quantities in $S^{(0)}$, we will follow \cite{Keresztes_2015, Bradley_2017, Tornkvist_2019,Bradley_2021} and make use of the spatial gradients of these quantities, as these gradients vanish on the background due to the assumed homogeneity.  We therefore introduce the following gauge invariant variables
\begin{alignat}{4}
 \tensor{X}{_a} &\equiv \tensor{\delta}{_a}\mathcal{E}, \quad  &\tensor{V}{_a} &\equiv \tensor{\delta}{_a}\Sigma, \quad &\tensor{W}{_a} &\equiv \tensor{\delta}{_a}\Theta, \quad &\tensor{Y}{_a} &\equiv \tensor{\delta}{_a}\Pi, \notag \\ 
 \tensor{\mu}{_a}  &\equiv \tensor{\delta}{_a}\mu, \quad & \tensor{p}{_a} & \equiv \tensor{\delta}{_a}p,  \quad & \tensor{\mathcal{B}}{_a} & \equiv \tensor{\delta}{_a}\mathfrak{B}, \quad &  \tensor{ {\mathcal{Z}_{(i)}} }{_a} &\equiv \tensor{\delta}{_a}{{\mathcal{N}_{(i)}}}, \\
 &  \quad &  \tensor{{\mu_{(i)}}}{_a}  &\equiv \tensor{\delta}{_a}{\mu_{(i)}}, \quad & \tensor{{p_{(i)}}}{_a} & \equiv \tensor{\delta}{_a}{p_{(i)}}. \quad  & & \notag
\end{alignat}
Note that we do not introduce any new gauge invariant variables using the "$\hat{\quad}$" gradients here, as these can be given in terms of the "$\delta$" gradients and the vorticity after performing the intended harmonic decompositions later on \cite{Tornkvist_2019, Bradley_2021}. Our complete set of gauge invariant first order variables therefore becomes
\begin{equation}
\begin{split}
S^{(1)}  \equiv \{ &\tensor{X}{_a}, \tensor{V}{_a}, \tensor{W}{_a},\tensor{\mu}{_a},\tensor{p}{_a},\tensor{Y}{_a}, \mathcal{A}, \tensor{\mathcal{A}}{_a}, \tensor{\Sigma}{_a}, \tensor{\Sigma}{_a_b}, \tensor{\mathcal{E}}{_a}, \tensor{\mathcal{E}}{_a_b}, \\
&\mathcal{H}, \tensor{\mathcal{H}}{_a}, \tensor{\mathcal{H}}{_a_b}, \tensor{\alpha}{_a}, \tensor{a}{_a}, \xi, \phi, \tensor{\zeta}{_a_b}, \Omega, \tensor{\Omega}{_a}, Q, \tensor{Q}{_a}, \tensor{\Pi}{_a}, \\
&\tensor{{\mu_{(i)}}}{_a}, \tensor{{p_{(i)}}}{_a},\tensor{ {\mathcal{Z}_{(i)}} }{_a},  {\mathcal{V}_{(i)}}, \tensor{{\mathcal{V}_{(i)}}}{_a}, {\varepsilon_{(i)}}, {\mathcal{F}_{(i)}},\tensor{{\mathcal{F}_{(i)}}}{^a}, \\
& \mathfrak{E}, \tensor{\mathfrak{E}}{_a},\tensor{\mathfrak{B}}{_a}, \tensor{\mathcal{B}}{_a}, \rho, \mathcal{J}, \tensor{\mathcal{J}}{_a} \}.
\end{split}
\end{equation}

To obtain linearized equations governing the variables in $S^{(1)}$, we write the decomposed equations from Sec.~\ref{sec:Spacetime dynamics} in terms of these quantities by using commutation relations for the projected derivatives, which can be found in appendix~\ref{sec:Commutation Relations}.  In doing so, we also omit every term that is of second order or higher. This yields a large set of equations involving derivatives of the projected variables with respect to both time and space. Since the equations originating from the Ricci and Bianchi identities can be found in \cite{Bradley_2021}, we will here only state the equations related to the electromagnetic fields, the individual plasma components, and the decomposition of the total energy-momentum quantities in terms of their constituents. 

\subsection{Decomposition of the Total Energy-Momentum Tensor}

\begin{align}
\tensor{\mu}{^a} &= \sum_{i}\tensor{{\mu} }{_{(i)}^a} + \mathfrak{B}\tensor{\mathcal{B}}{^a}, \\
~
\tensor{p}{^a} &= \sum_{i}\tensor{p }{_{(i)}^a} + \frac{\mathfrak{B}}{3}\tensor{\mathcal{B}}{^a}, \\
~
Q &= \sum_{i}\left({\mu_{(i)}}  + {p_{(i)}} \right){\mathcal{V}_{(i)}} , \\
~
\tensor{Q}{^a} &= \sum_{i}\left({\mu_{(i)}}  + {p_{(i)}} \right)\tensor{{\mathcal{V}_{(i)}} }{^a} + \tensor{\epsilon}{^a^b}\mathfrak{B}\tensor{\mathfrak{E}}{_b}, \\
~
\tensor{Y}{^a} &= -\frac{4\mathfrak{B}}{3}\tensor{\mathcal{B}}{^a}, \\
~
\tensor{\Pi}{^a} &= -\mathfrak{B}\tensor{\mathfrak{B}}{^a}.
\end{align}

\subsection{Equations for Each Fluid Component}

\begin{align}
\begin{split}
 {\dot{\mu}} \indices{_{(i)}^{\bar{a}}} = &~\tensor{\mathcal{A}}{^a}{\dot{\mu}_{(i)}}  + \frac{1}{2}\left(\Sigma - \frac{2\Theta}{3}\right)\tensor{\mu}{_{(i)}^a} - \left(\tensor{\mu}{_{(i)}^a} +\tensor{p}{_{(i)}^a}\right)\Theta\\
~
& - \left({\mu_{(i)}} +{p_{(i)}} \right)\left(\tensor{\delta}{^a}\hat{\mathcal{V}}_{(i)}  + \tensor{\delta}{^a}\tensor{\delta}{^b}\tensor{{\mathcal{V}} }{_{(i)}_b} + \tensor{W}{^a}\right) + \tensor{\delta}{^a}{\varepsilon_{(i)}} ,
\end{split} \\
~
\begin{split}
\left({\mu_{(i)}}  + {p_{(i)}} \right)\tensor{\delta}{^a}{\dot{\mathcal{V}}_{(i)}}  = &-\tensor{{\hat{p}} }{_{(i)}^{\bar{a}}}+\left( 2\tensor{\epsilon}{^a^b}\tensor{\Omega}{_b}-\tensor{\delta}{^a}{\mathcal{V}_{(i)}} \right){\dot{p}_{(i)}}   + {\rho_{(i)}}\tensor{\delta}{^a}\mathfrak{E} \\
~
&- \left({\mu_{(i)}}  + {p_{(i)}} \right)\left( \tensor{\delta}{^a}\mathcal{A} + \left(\Sigma + \frac{\Theta}{3}\right)\tensor{\delta}{^a}{\mathcal{V}_{(i)}} \right)+ \tensor{\delta}{^a}{\mathcal{F}_{(i)}}, 
\end{split}\\
~
\begin{split}
\left({\mu_{(i)}}  + {p_{(i)}} \right)\tensor{{\dot{\mathcal{V}}} }{_{(i)}^{\bar{a}}} = &-{\dot{p}_{(i)}} \tensor{{\mathcal{V}_{(i)}} }{^a} - \tensor{{p_{(i)}} }{^a} + {\rho_{(i)}}\left( \tensor{\mathfrak{E}}{^a} + \mathfrak{B}\tensor{\epsilon}{^a^b}\tensor{\mathcal{V} }{_{(i)}_b}\right) \\
~
&-\left({\mu_{(i)}}  + {p_{(i)}} \right)\left( \tensor{\mathcal{A}}{^a} - \frac{1}{2}\left(\Sigma-\frac{2\Theta}{3} \right)\tensor{{\mathcal{V}_{(i)}} }{^a} \right) + \tensor{\mathcal{F}}{_{(i)}^a}, 
\end{split}\\
~
\begin{split}
\tensor{{\dot{\mathcal{Z}}} }{_{(i)}^{\bar{a}}} = &~\tensor{\mathcal{A}}{^a}{\dot{\mathcal{N}}_{(i)}}  + \left(\frac{\Sigma}{2} - \frac{4\Theta}{3}\right)\tensor{{\mathcal{Z}} }{_{(i)}^a}  \\
~
&- {\mathcal{N}_{(i)}} \left(\tensor{W}{^a} + \tensor{\delta}{^a}{\hat{\mathcal{V}}_{(i)}}  + \tensor{\delta}{^a}\tensor{\delta}{^b}\tensor{\mathcal{V} }{_{(i)}_b}\right). 
\end{split}
\end{align}

\subsection{Maxwell's Equations}

\begin{align}
\begin{split}
\dot{\mathfrak{E}} - \tensor{\epsilon}{_a_b}\tensor{\delta}{^a}\tensor{\mathfrak{B}}{^b} =&~ 2\xi\mathfrak{B} + \left(\Sigma - \frac{2\Theta}{3}\right)\mathfrak{E} -\mathcal{J}, 
\end{split}\\
~
\begin{split}
\hat{\mathfrak{E}} + \tensor{\delta}{_a}\tensor{\mathfrak{E}}{^a} =&~ 2\mathfrak{B}\Omega + \rho,
\end{split}\\
~
\begin{split}
 \tensor{\dot{\mathcal{B}}}{^{\bar{a}}} + \tensor{\epsilon}{_b_c}\tensor{\delta}{^a}\tensor{\delta}{^b}\tensor{\mathfrak{E}}{^c} =&~ \tensor{\mathcal{A}}{^a}\dot{\mathfrak{B}} + \frac{3}{2}\left(\Sigma - \frac{2\Theta}{3}\right)\tensor{\mathcal{B}}{^a}  + \left(\tensor{V}{^a} - \frac{2}{3}\tensor{W}{^a}\right)\mathfrak{B},
\end{split}\\
~
\begin{split}
\tensor{\hat{\mathcal{B}}}{^{\bar{a}}} + \tensor{\delta}{^a}\tensor{\delta}{^b}\tensor{\mathfrak{B}}{_b} = &~2\tensor{\epsilon}{^a^b}\tensor{\Omega}{_b}\dot{\mathfrak{B}} - \mathfrak{B}\tensor{\delta}{^a}\phi,
\end{split}\\
~
\begin{split}
\tensor{\dot{\mathfrak{E}}}{_{\bar{a}}} + \tensor{\epsilon}{_a_b}\left(\tensor{\hat{\mathfrak{B}}}{^b} - \tensor{\mathcal{B}}{^b}\right) = & -\frac{1}{2}\left(\Sigma + \frac{4\Theta}{3}\right)\tensor{\mathfrak{E}}{_a} + \mathfrak{B}\tensor{\epsilon}{_a_b}\left(\tensor{\mathcal{A}}{^b} - \tensor{a}{^b}\right)  - \tensor{\mathcal{J}}{_a},
\end{split}\\
~
\begin{split}
\tensor{\dot{\mathfrak{B}}}{_{\bar{a}}} - \tensor{\epsilon}{_a_b}\left(\tensor{\hat{\mathfrak{E}}}{^b} - \tensor{\delta}{^b}\mathfrak{E}\right) = &-\frac{1}{2}\left(\Sigma + \frac{4\Theta}{3}\right)\tensor{\mathfrak{B}}{_a}  + \mathfrak{B}\left(-\tensor{\alpha}{_a} +\tensor{\Sigma}{_a} + \tensor{\epsilon}{_a_b}\tensor{\Omega}{^b}\right), 
\end{split}\\
~
\begin{split}
\tensor{\delta}{^a}\rho = &~\sum_{i} q_{c(i)}\tensor{{\mathcal{Z}_{(i)}} }{^a},
\end{split}\\
~
\begin{split}
\mathcal{J} = &~\sum_{i} q_{c(i)}{\mathcal{N}_{(i)}} {\mathcal{V}_{(i)}} ,
\end{split}\\
~
\begin{split}
\tensor{\mathcal{J}}{_a} =&~ \sum_{i} q_{c(i)}{\mathcal{N}_{(i)}} \tensor{{\mathcal{V}_{(i)}} }{_a}.
\end{split}
\end{align}

\section{Harmonic Decomposition}

To simplify the linearized equations, they are then harmonically decomposed using harmonics $P^{k_{\parallel}}(z)$ and  $Q^{k_{\perp}}(\vartheta,\varphi)$, which are suited for homogeneous and hypersurface orthogonal LRS class II backgrounds with $\phi = 0$ \cite{Bradley_2017}. These harmonics are defined on the background as functions satisfying the equations
\begin{alignat}{3}
\hat{\Delta}P^{k_{\parallel}} & = -\frac{k_{\parallel}^2}{a_1^2}P^{k_{\parallel}} , \quad &\tensor{\delta}{_a}P^{k_{\parallel}}  &= 0, \quad &\dot{P}^{k_{\parallel}} &= 0, \\
     \delta^2 Q^{k_{\perp}} & = -\frac{k_{\perp}^2}{a_2^2}Q^{k_{\perp}}, \quad & \hat{Q}^{k_{\perp}} &= 0, \quad & \dot{Q}^{k_{\perp}} &= 0,
\end{alignat}
where the differential operators are defined as $\hat{\Delta} \equiv \tensor{n}{^a}\tensor{\nabla}{_a}\tensor{n}{^b}\tensor{\nabla}{_b}$, and $\delta^2 = \tensor{\delta}{_a}\tensor{\delta}{^a} $, whilst $k_{\parallel}$ and $k_{\perp}$ are comoving dimensionless wave numbers. Using these harmonics, scalars can be decomposed as 
\begin{equation}
\Psi = \sum\limits_{k_{\parallel}, k_{\perp}} \Psi ^S_{k_{\parallel} k_{\perp}}P^{k_{\parallel}}Q^{k_{\perp}}, 
\end{equation}
where the coefficients $\Psi ^S$ now only depend on time. To decompose 2-vectors $\tensor{\Psi}{_a}$ and PSTF 2-tensors $\tensor{\Psi}{_a_b} $, we also introduce vector and tensor harmonics following 
\begin{alignat}{2}
\tensor{{Q^{k_{\perp}}}}{_a} &= a_2\tensor{\delta}{_a}Q^{k_{\perp}}, \quad & \tensor{{Q^{k_{\perp}}}}{_a_b} &= a_2^2\tensor{\delta}{_{\{a}}\tensor{\delta}{_{b\}}}Q^{k_{\perp}},\\
 \tensor{{\overline{Q}^{k_{\perp}}}}{_a} &= a_2\tensor{\epsilon}{_a_b}\tensor{\delta}{^b}Q^{k_{\perp}},  \quad & \tensor{{\overline{Q}^{k_{\perp}}}}{_a_b} &= a_2^2\tensor{\epsilon}{_c_{\{a}}\tensor{\delta}{^c}\tensor{\delta}{_{b\}}}Q^{k_{\perp}}.
\end{alignat}
Using these harmonics, we may write
\begin{align}
\tensor{\Psi}{_a} &= \sum\limits_{k_{\parallel}, k_{\perp}} P^{k_{\parallel}}\left(\Psi ^V_{k_{\parallel} k_{\perp}}\tensor{{Q^{k_{\perp}}} }{_a} +  \overline{\Psi} ^V_{k_{\parallel} k_{\perp}}\tensor{{\overline{Q}^{k_{\perp}}}}{_a}\right), \\
\tensor{\Psi}{_a_b} &= \sum\limits_{k_{\parallel}, k_{\perp}} P^{k_{\parallel}}\left(\Psi ^T_{k_{\parallel} k_{\perp}}\tensor{{Q^{k_{\perp}}}}{_a_b}+  \overline{\Psi} ^T_{k_{\parallel} k_{\perp}}\tensor{{\overline{Q}^{k_{\perp}}}}{_a_b}\right).
\end{align}
Here we have used bars to denote the so called odd parts of the decompositions, whilst the even parts are written without bars. In the following, when considering the relations for the individual harmonic coefficients, we will omit their subscripts, as this should not lead to any ambiguities.

Inserting the harmonic decompositions into the linearized equations and using the properties of the harmonics, described in appendix \ref{sec:Properties of the Harmonics}, the system can be reduced significantly by solving for some of the harmonic coefficients in terms of the others. The harmonic decompositions of the linearized equations, which were presented in Sec.~\ref{sec:Perturbations}, can be found in appendix \ref{sec:Harmonic Coefficients in the Multifluid System}. However, here we encounter a difficulty which was not present in previous papers. At this point in previous works, the system effectively divides into two separate parts, one containing the odd coefficients and another containing the even parts \cite{Bradley_2017, Tornkvist_2019, Bradley_2021}. One exception to the division into even and odd equations, which is also present in previous works, are quantities defined using a Levi-Civita psuedo-tensor, such as the vorticity and the magnetic part of the Weyl tensor. For these quantities it is generally the even coefficients that are present in the odd system and vice versa, but the systems still decouple from each other. However, when including a non-zero magnetic field to zeroth order, the division into even and odd subsystems does not seem possible in general, as the pseudoscalar $\mathfrak{B}$ serves as a coupling that connects the systems together. This can be seen when considering the harmonically decomposed momentum equations for the fluid components 
\begin{align}
\begin{split}
\left(\mu_{(i)} + p_{(i)}\right)\dot{\mathcal{V}}\indices{_{(i)}^V} = &- \dot{p}_{(i)} \mathcal{V}\indices{_{(i)}^V }- p\indices{_{(i)}^V} + \rho_{(i)}\left( \mathfrak{E}^V - \mathfrak{B}\overline{\mathcal{V}}\indices{_{(i)}^V}\right) \\
~
&-\left(\mu_{(i)} + p_{(i)}\right)\left( \mathcal{A}^V - \frac{1}{2}\left(\Sigma-\frac{2\Theta}{3} \right)\mathcal{V}\indices{_{(i)}^V }\right) + \mathcal{F}\indices{_{(i)}^V },
\end{split} \label{eq:Vidot}\\
~ 
\begin{split}
\left(\mu_{(i)} + p_{(i)}\right)\dot{\overline{\mathcal{V}} }\indices{_{(i)}^V}  = &-\dot{p}_{(i)} \overline{\mathcal{V}}\indices{_{(i)}^V } - \overline{p}\indices{_{(i)}^V}  + \rho_{(i)}\left( \overline{\mathfrak{E}}^V + \mathfrak{B} \mathcal{V}\indices{_{(i)}^V} \right) \\
~
&-\left(\mu_{(i)} + p_{(i)}\right)\left( \overline{\mathcal{A}}^V - \frac{1}{2}\left(\Sigma-\frac{2\Theta}{3} \right)\overline{\mathcal{V}}\indices{_{(i)}^V } \right) +  \overline{\mathcal{F}}\indices{_{(i)}^V }.
\end{split} \label{eq:Vibardot}
\end{align}
Here it can be explicitly observed that $\mathcal{V}\indices{_{(i)}^V }$ is coupled to $\overline{\mathcal{V}}\indices{_{(i)}^V }$ through the terms proportional to $\mathfrak{B}$. To avoid this problem and simplify matters even further, we will now apply the cold MHD approximation

\section{Cold MHD Approximation}

In the cold MHD approximation, we assume that  $\mu_{(i)} \approx m_{(i)}\mathcal{N}_{(i)}$ and $p_{(i)}\approx 0$ to both zeroth and first order, and write the harmonic equations in terms of the collective variables
\begin{alignat}{2}
\mu_{(F)} & \equiv \sum_{i}\mu_{(i)}, \quad  & \mu_{(F)}\mathcal{V}\indices{_{(F)}^S } &\equiv \sum_{i}\mu_{(i)}\mathcal{V}\indices{_{(i)}^S}, \notag \\
\mu\indices{_{(F)}^V} &\equiv \sum_{i}\mu\indices{_{(i)}^V}, \quad  & \mu_{(F)}\mathcal{V}\indices{_{(F)}^V } &\equiv \sum_{i}\mu_{(i)}\mathcal{V}\indices{_{(i)}^V}, \\
\overline{\mu}\indices{_{(F)}^V} &\equiv \sum_{i}\overline{\mu}\indices{_{(i)}^V}, \quad  & \mu_{(F)}\overline{\mathcal{V}}\indices{_{(F)}^V } &\equiv \sum_{i}\mu_{(i)}\overline{\mathcal{V}}\indices{_{(i)}^V} \notag.
\end{alignat}
The equations governing these variables are found by adding the equations for the individual fluid components. Using charge neutrality to zeroth order and assuming that the plasma consists of electrons and a much more massive ion counterpart, it can then be shown that the collective velocities satisfy the equations
\begin{align}
\mu_{(F)} \dot{\mathcal{V}}\indices{_{(F)}^V} =&  -\frac{\mathfrak{B}}{\eta}\left(\overline{\mathfrak{E}}^V + \mathfrak{B}\mathcal{V}\indices{_{(F)}^V}\right)-\mu_{(F)}\left(\mathcal{A}^V - \frac{1}{2}\left(\Sigma-\frac{2\Theta}{3} \right)\mathcal{V}\indices{_{(F)}^V}\right), \label{eq:VFdot} \\
~ 
\mu_{(F)}\dot{\overline{\mathcal{V}}}\indices{_{(F)}^V} =&~ \frac{\mathfrak{B}}{\eta}\left(\mathfrak{E}^V - \mathfrak{B}\overline{\mathcal{V}}\indices{_{(F)}^V}\right)-\mu_{(F)}\left(\overline{\mathcal{A}}^V - \frac{1}{2}\left(\Sigma-\frac{2\Theta}{3} \right)\overline{\mathcal{V}}\indices{_{(F)}^V}\right), \label{eq:VFbardot}
\end{align}
where we have written the currents as 
\begin{align}
\mathcal{J}^V &= \frac{1}{\eta}\left(\mathfrak{E}^V - \mathfrak{B}\overline{\mathcal{V}}\indices{_{(F)}^V}\right), \label{eq:Current1} \\
~
\overline{\mathcal{J}}^V  &= \frac{1}{\eta}\left(\overline{\mathfrak{E}}^V + \mathfrak{B}\mathcal{V}\indices{_{(F)}^V}\right), \label{eq:Current2}
\end{align}
using the linearised and harmonically decomposed version of Eq.~(\ref{eq:OhmsLaw}). The other equations obtained on rewriting the multifluid system in this manner can be found in appendix \ref{sec:Harmonic Coefficients in the Cold MHD System}. The momentum equations, Eqs.~(\ref{eq:VFdot})–(\ref{eq:VFbardot}), should be compared to Eqs.~(\ref{eq:Vidot})–(\ref{eq:Vibardot}). At the cost of introducing an electrical resistivity $\eta$, we see that we no longer have any coupling between the even and odd velocities in Eqs.~(\ref{eq:VFdot})–(\ref{eq:VFbardot}), in contrast to Eqs.~(\ref{eq:Vidot})–(\ref{eq:Vibardot}). This simplifies matters, and the system is ultimately seen to decouple into two subsectors, as shown in Sec.~\ref{sec:Final System}.  However, we will need some model for the electrical resistivity if we want to completely close the system. Therefore, when performing numerical calculations in Sec.~\ref{sec:Analyzing the Final System}, we will make the simplified assumption that $\eta$ follows a Spitzer-like expression so that 
\begin{equation}
\eta \propto {T_{(e)}}^{-3/2},
\label{eq:Spitzer}
\end{equation}
where $T_{(e)}$ is the electron temperature \cite{Baumjohann_2012}. Assuming that the electron pressure, although neglected in the main equations, follows a polytropic equation of state and an ideal gas law,
\begin{equation}
p_{(e)} \propto {\mathcal{N}_{(e)}}^{\gamma} \propto \mathcal{N}_{(e)}T_{(e)},
\end{equation}
it follows from the electron number conservation and Eq.~(\ref{eq:Spitzer}) that 
\begin{equation}
\dot{\eta}  = \frac{3\left(\gamma-1\right)}{2}\Theta \eta.  \label{eq:etadot}
\end{equation}

\section{Final System}\label{sec:Final System}

The harmonically decomposed system of equations in the cold MHD approximation can be significantly reduced by solving for some of the coefficients in terms of the others. However, before arriving at the final system, we should note that there is still some freedom in specifying the dyad $\{ u^a, n^a \}$ to first order. To zeroth order, this dyad is fixed. The vector field $u^a$ is then assumed to be orthogonal to the hypersurfaces of homogeneity, and coincides with the plasma velocities. As for the field $n^a$, this is fixed since it defines the direction of the local rotational symmetry. However, when going to first order, the homogeneity and rotational symmetry are broken, and the vector field $u^a$ does not need to coincide with the plasma velocities. As a result, there is some freedom in defining $u^a$ and  $n^a$ to first order. We will use this freedom to set $a^V$, $\overline{a}^V$, $\mathcal{A}^S$, $\mathcal{A}^V$, and  $\overline{\mathcal{A}}^V $ to zero. For details on the effect and use of infinitesimal transformations of the dyad $\{ u^a, n^a \}$, the reader is referred to \cite{Keresztes_2015}.

Finally, after specifying the dyad to first order, we arrive at two subsystems for the first order quantities. These systems are seen to close independently of each other and consist of the odd sector $\{\Omega^S, \overline{\mathcal{E}}^T, \mathcal{H}^T, {\overline{\mathcal{V}}_{(F)}}^V, \mathfrak{E}^S, \mathfrak{E}^V, \overline{\mathfrak{B}}^V\}$, and the even sector $\{\overline{\Omega}^V, \overline{\mathcal{H}}^T, \mathcal{E}^T, \Sigma^T, {\mu_{(F)}}^V$, $ {\mathcal{V}_{(F)}}^S, {\mathcal{V}_{(F)}}^V, \mathcal{B}^V, \overline{\mathfrak{E}}^V \}$. All other coefficients can be given algebraically in terms of these sets. 
\subsection{Odd Sector}
The coefficients in the odd sector satisfy the following evolution equations
\begin{align}
	\dot{\Omega}^S =&~\left(\Sigma - \frac{2\Theta}{3}\right)\Omega^S,  \label{eq:OddODEFirst} \\
	~ 
	\begin{split}
	\dot{\overline{\mathcal{E}}}^T =& -\frac{3}{2}\left( F + \Sigma D \right) \overline{\mathcal{E}}^T + \frac{i\pk}{a_1}\left(1-D\right)\mathcal{H}^T  \\
	&+ P\Omega^S + \frac{1}{a_2}\left(1+D\right)\left({\mu_{(F)}}{\overline{\mathcal{V}}_{(F)}}^V + \mathfrak{B}\mathfrak{E}^V\right), 
	\end{split}  \\
	~ 
	\begin{split}
	\dot{\mathcal{H}}^T =&~\left(\frac{3ia_1\Pi}{\pk\okk}\left(\Sigma + \frac{\Theta}{3} \right) + S \right)\Omega^S + \frac{1}{a_2}\mathfrak{B}\overline{\mathfrak{B}}^V  \\	
	~
	&+ \frac{ia_1}{a_2\pk B}\Bigg[\left( \Sigma+ \frac{\Theta}{3}\right)\left(3\Pi + \frac{\okk}{a_2^2}\right)   \\
	~ 
	&-\frac{\pkk}{a_1^2}\left(\Sigma - \frac{2\Theta}{3}\right) \Bigg] \left({\mu_{(F)}}{\overline{\mathcal{V}}_{(F)}}^V + \mathfrak{B}\mathfrak{E}^V\right) \\
	~
	 &+\frac{ia_1}{2\pk}\left( \frac{2\pkk}{a_1^2} - CB + 9\Sigma E + 3\Pi\right) \overline{\mathcal{E}}^T - \frac{3}{2}\left(2E + F\right) \mathcal{H}^T,  
	\end{split} \\
~
	\begin{split}
	{\dot{\overline{\mathcal{V}}}_{(F)}}^V =&~\frac{\mathfrak{B}}{\eta{\mu_{(F)}}}\left(\mathfrak{E}^V - \mathfrak{B}{\overline{\mathcal{V}}_{(F)}}^V\right)+ \frac{1}{2}\left(\Sigma-\frac{2\Theta}{3} \right){\overline{\mathcal{V}}_{(F)}}^V, 
	\end{split}\label{eq:VFbarV} \\
	~ 
	\begin{split}
	\dot{\mathfrak{E}}^S  =&~\frac{\okk}{a_2}\overline{\mathfrak{B}}^V -\frac{2ia_1}{\pk}\mathfrak{B} \left( \Sigma + \frac{\Theta}{3} \right)\Omega^S  + \left(\Sigma - \frac{2\Theta}{3} - \frac{1}{\eta}\right)\mathfrak{E}^S, 
	\end{split} \label{ps:eq19} \\
	~ 
	\begin{split}
	\dot{\mathfrak{E}}^V = &~\frac{i\pk}{a_1}\overline{\mathfrak{B}}^V -\frac{1}{2}\left(\Sigma + \frac{4\Theta}{3}\right)\mathfrak{E}^V  -\frac{1}{\eta}\left(\mathfrak{E}^V - \mathfrak{B}{\overline{\mathcal{V}}_{(F)}}^V\right)- \frac{2a_2\dot{\mathfrak{B}}}{k_{\bot}^2}\Omega^S,
 	\end{split} \label{ps:eq56} \\
	~
	\begin{split}
	\dot{\overline{\mathfrak{B}}}^V  = &~\frac{i\pk}{a_1}\mathfrak{E}^V - \frac{1}{a_2}\mathfrak{E}^S-\frac{1}{2}\left(\Sigma + \frac{4\Theta}{3}\right)\overline{\mathfrak{B}}^V  \\
	~
	& -\frac{ia_1}{\pk} \mathfrak{B}\Bigg[ \frac{2\left(\mathcal{R}a_2^2 - \okk\right)}{a_2B} \left ( \frac{3\Sigma}{2} \overline{\mathcal{E}}^T  + \frac{i\pk}{a_1} \mathcal{H}^T \right ) \\
~ 
	&- \frac{2}{B} \left( 3\Sigma\left (\Sigma + \frac{\Theta}{3}\right ) + \frac{2\pkk}{a_1^2}  \right )\left({\mu_{(F)}}{\overline{\mathcal{V}}_{(F)}}^V  +  \mathfrak{B}\mathfrak{E}^V\right) \\
	&  -\frac{a_2}{\okk} \left ( \mu + p - \frac{\Pi}{2} -3\mathcal{E}+ 2N + 3\Sigma \left ( \Sigma + \frac{\Theta}{3} \right )  +\frac{2\pkk}{a_1^2} \right )\Omega^S \Bigg], 
	\end{split} \label{eq:OddODELast}
\end{align}
where the auxiliary variables are defined as
\begin{align}
	\tilde{k}^2 =&~ \frac{\okk}{a_2^2} + \frac{2\pkk}{a_1^2},\\
	~ 
	B=&~ \tilde{k}^2 + \frac{9}{2}\Sigma^2 + 3\left(\mathcal{E}+ \frac{1}{2}\Pi \right),  \label{eq:auxB}\\
	~ 
	CB = &~\Sigma \left( \Theta - \frac{3\Sigma}{2} \right) - \frac{\okk}{a_2^2},  \\
	~ 
	F = &~\Sigma + \frac{2\Theta}{3},  \\
	~ 
	D =&~ C + \frac{\mu + p - 2\Pi}{B},   \\
	~
	P = &~\frac{2}{\okk B} \left( \mu + p -\frac{1}{2}\Pi \right)\left( B + 3\mathcal{E}+ \mu + p - \frac{1}{2}\Pi +\mathcal{R} - \frac{\okk}{a_2^2} \right), \\
	~ 
	S = &~ \frac{2ia_1}{3\pk\okk B}\left( \mu + p - \frac{1}{2}\Pi\right) \left( 3\Sigma\left( \frac{\okk}{a_2^2} - \frac{\pkk}{a_1^2} + 3\Pi\right) + \Theta\left( \tilde{k}^2 + 3\Pi \right) \right), \\
	~ 
	EB = &~ \frac{\Sigma}{2}\left(CB - \mathcal{E}- \frac{5}{2}\Pi \right) + \frac{\Theta}{3}\left(\mathcal{E}- \frac{1}{2}\Pi \right), \\
	~ 
	N = &~ \left( \mu + p - \frac{1}{2}\Pi \right) \left( 1+ \frac{2}{a_2^2 B}\left(\mathcal{R}a_2^2 - \okk \right) \right).
\end{align}

We get the remaining relevant harmonic coefficients as algebraic expressions in terms of $\Omega^S, \overline{\mathcal{E}}^T, \mathcal{H}^T, {\overline{\mathcal{V}}_{(F)}}^V, \mathfrak{E}^S, \mathfrak{E}^V$, and $\overline{\mathfrak{B}}^V$. The remaining scalar coefficients are
\begin{align}
\rho^S =&~ \frac{i\pk}{a_1}\mathfrak{E}^S -\frac{\okk}{a_2}\mathfrak{E}^V - 2\mathfrak{B}\Omega^S, \quad \frac{ i\pk}{a_1}\xi^S =~ \left( \Sigma + \frac{\Theta}{3} \right)\Omega^S, \\
~
\begin{split}
    \frac{ia_2\pk}{a_1\okk}\mathcal{H}^S =  &~\frac{ \left (\mathcal{R}a_2^2 - \okk \right )}{a_2B} \left ( \frac{3\Sigma}{2} \overline{\mathcal{E}}^T  + \frac{i\pk}{a_1}\mathcal{H}^T \right )  -\frac{a_2}{\okk}\left ( \mu + p - \frac{\Pi}{2} + N \right )\Omega^S \\ 
    & -\frac{1}{B}\left ( 3\Sigma\left (\Sigma + \frac{\Theta}{3} \right )  + \frac{2\pkk}{a_1^2} \right )\left(\mu_{(F)}\overline{\mathcal{V}}\indices{_{(F)}^V}  +  \mathfrak{B}\mathfrak{E}^V\right).
\end{split} 
\end{align}
The remaining vector coefficients are in turn given by
\begin{align}
 \Omega^V =&~ \frac{ia_2\pk}{a_1k_{\bot}^2}\Omega^S, \\
~
 \overline{Q}^V =&~ \mu_{(F)}\overline{\mathcal{V}}\indices{_{(F)}^V}  +  \mathfrak{B}\mathfrak{E}^V, \\
~
\overline{\mu}\indices{_{(F)}^V} =&~ \frac{2a_2\dot{\mu}_{(F)}}{k_{\bot}^2}\Omega^S, \quad  \overline{\mathcal{B}}^V =~ \frac{2a_2\dot{\mathfrak{B}}}{k_{\bot}^2}\Omega^S, \quad   \overline{\mu}^V =~ \overline{\mu}\indices{_{(F)}^V} + \mathfrak{B}\overline{\mathcal{B}}^V, \\
~
\overline{W}^V =&~ \frac{2a_2\dot{\Theta}}{k_{\bot}^2}\Omega^S, \quad  \overline{X}^V =~ \frac{2a_2\dot{\mathcal{E}}}{k_{\bot}^2}\Omega^S, \quad  \overline{V}^V =~ \frac{2a_2\dot{\Sigma}}{k_{\bot}^2}\Omega^S, \\
~
\overline{p}^V =&~  \frac{\mathfrak{B}}{3}\overline{\mathcal{B}}^V, \quad  \overline{Y}^V = -\frac{4\mathfrak{B}}{3}\overline{\mathcal{B}}^V, \quad \overline{\Pi}^V = -\mathfrak{B}\overline{\mathfrak{B}}^V, \\
~
 \begin{split}
     \frac{i\pk}{a_1}\overline{\Sigma}^V = &-\frac{a_2}{\okk} \left ( \mu + p - \frac{\Pi}{2} + N + 3\Sigma \left ( \Sigma + \frac{\Theta}{3} \right )  +\frac{\pkk}{a_1^2} \right )\Omega^S \\ 
     & + \frac{(\mathcal{R}a_2^2-\okk)}{a_2B} \left ( \frac{3\Sigma}{2}\overline{\mathcal{E}}^T  + \frac{i\pk}{a_1}\mathcal{H}^T \right ) \\
     & - \frac{1}{B} \left( 3\Sigma \left ( \Sigma + \frac{\Theta}{3} \right ) + \frac{2\pkk}{a_1^2} \right )\left(\mu_{(F)}\overline{\mathcal{V}}\indices{_{(F)}^V}  +  \mathfrak{B}\mathfrak{E}^V\right),
\end{split}\\
~
\begin{split}
     \overline{\mathcal{E}}^V = &~  \frac{ia_1a_2}{\pk\okk} \Bigg( \left ( 3\mathcal{E} + \frac{\Pi}{2} +2\mu+2p\right)\left (\Sigma + \frac{\Theta}{3}\right )  \\ 
     &  -\frac{3\Sigma}{B}\left ( \mu + p - \frac{\Pi}{2} \right ) \left(\frac{2\pkk}{a_1^2} + 3\Sigma\left(\Sigma+\frac{\Theta}{3}\right)  \right) \Bigg) \Omega^S  \\
      &- \frac{\mathcal{R}a_2^2 - \okk}{2a_2}\left ( \frac{ia_1}{\pk} \left (1 - \frac{9\Sigma^2}{2B} \right)\overline{\mathcal{E}}^T   + \frac{3\Sigma}{B} \mathcal{H}^T  \right) + \frac{\mathfrak{B}}{2}\overline{\mathfrak{B}}^V \\
      &+\frac{ia_1}{\pk B} \Bigg[   \left( \Sigma + \frac{\Theta}{3}\right)\left (\tilde{k}^2 - \mathcal{R}  -\frac{3\Sigma}{2}\left(\Sigma - \frac{2\Theta}{3}\right) \right)   \\
      ~ 
      & - \frac{3\Sigma\pkk}{a_1^2}  \Bigg ]\left(\mu_{(F)}\overline{\mathcal{V}}\indices{_{(F)}^V}  +  \mathfrak{B}\mathfrak{E}^V\right), 
\end{split} \\
~
\begin{split}
        \mathcal{H}^V = &~\frac{a_2}{\okk}\left( 3\mathcal{E} - N\right )\Omega^S + \frac{\mathcal{R}a_2^2 - \okk}{a_2B} \left( \frac{3\Sigma}{2}\overline{\mathcal{E}}^T  + \frac{i\pk}{a_1}\mathcal{H}^T \right) \\
       & - \frac{1}{B}\left ( 3\Sigma\left (\Sigma + \frac{\Theta}{3} \right ) + \frac{2\pkk}{a_1^2} - \frac{B}{2} \right ) \left(\mu_{(F)}\overline{\mathcal{V}}\indices{_{(F)}^V}  +  \mathfrak{B}\mathfrak{E}^V\right),
\end{split}\\
~
\begin{split}
    \frac{i\pk}{a_1} \overline{\alpha}^V = &-\frac{a_2}{\okk} \left(3\mathcal{E}-N \right)\Omega^S - \frac{\mathcal{R}a_2^2 - \okk}{a_2B} \left ( \frac{3\Sigma}{2} \overline{\mathcal{E}}^T  + \frac{i\pk}{a_1} \mathcal{H}^T \right ) \\
    & + \frac{1}{B} \left( 3\Sigma\left (\Sigma + \frac{\Theta}{3}\right ) + \frac{2\pkk}{a_1^2}  \right )\left(\mu_{(F)}\overline{\mathcal{V}}\indices{_{(F)}^V}  +  \mathfrak{B}\mathfrak{E}^V\right). 
    \end{split}
\end{align}
Finally, the algebraic tensor coefficients are
\begin{align}
\begin{split}
    \frac{B}{2} \overline{\Sigma}^T = &- \frac{2}{\okk} \left(\mu + p + \frac{B-\Pi}{2} \right )\Omega^S + \frac{3\Sigma}{2}\overline{\mathcal{E}}^T  + \frac{i\pk}{a_1}\mathcal{H}^T \\
    ~ 
    &- \frac{1}{a_2}\left(\mu_{(F)}\overline{\mathcal{V}}\indices{_{(F)}^V}  +  \mathfrak{B}\mathfrak{E}^V\right),
 \end{split}\label{eq:barSigmaT} \\
 ~ 
\begin{split}
        \frac{B}{2}\overline{\zeta}^T = &~ \frac{2ia_1}{\pk\okk} \left ( \Sigma + \frac{\Theta}{3} \right )\left ( \mu + p + \frac{B-\Pi}{2} \right )\Omega^S - \frac{ia_1}{2\pk}\left ( \mathcal{R} - \tilde{k}^2 \right ) \overline{\mathcal{E}}^T \\
        &+\left ( \Sigma + \frac{\Theta}{3} \right )\mathcal{H}^T + \frac{ia_1}{a_2\pk}\left (\Sigma + \frac{\Theta}{3} \right )\left(\mu_{(F)}\overline{\mathcal{V}}\indices{_{(F)}^V}  +  \mathfrak{B}\mathfrak{E}^V\right).
\end{split}
\end{align}


\subsection{Even Sector}

The evolution equations for the coefficients in the even sector are
\begin{align}
		\begin{split}
		\dot{\mu}\indices{_{(F)}^V} = &~\left(\frac{\Sigma}{2} - \frac{4\Theta}{3} - \frac{3\Sigma{\mu_{(F)}}}{2B}\right){\mu_{(F)}}^V - \frac{3\Sigma{\mu_{(F)}}\mathfrak{B}}{2B}\mathcal{B}^V \\
		~
		& + \frac{ia_1{\mu_{(F)}}}{2a_2\pk}\left( \left( 1- \frac{3\Sigma}{B}\left(\Sigma + \frac{\Theta}{3}\right)\right){\mu_{(F)}} - \frac{2\pkk}{a_1^2}\right){\mathcal{V}_{(F)}}^S \\
		~ 
		&-{\mu_{(F)}}\left( \left( 1 +\frac{3\Sigma}{2B}\left(\Sigma - \frac{2\Theta}{3}\right)\right) {\mu_{(F)}}-\frac{\okk}{a_2^2}\right){\mathcal{V}_{(F)}}^V \\
		~ 
		&+\frac{3a_2\Sigma{\mu_{(F)}}}{2}\left(C-1\right)\mathcal{E}^T - \frac{a_2B{\mu_{(F)}}}{2}\left(C-1\right)\Sigma^T \\
		~
		&-\frac{ia_2\pk{\mu_{(F)}}}{2a_1}\left(J-2\right)\overline{\mathcal{H}}^T - \frac{ia_1{\mu_{(F)}}}{\pk}\left( \frac{G}{B} + \frac{2\pkk}{a_1^2}\right)\overline{\Omega}^V \\
		~
		&+ \frac{\mathfrak{B}{\mu_{(F)}}}{2B}\left( 2B + 3\Sigma\left(\Sigma-\frac{2\Theta}{3}\right)\right)\overline{\mathfrak{E}}^V,
	\end{split} \label{eq:EvenODEFirst}\\
	~ 
	\begin{split}
		\dot{\mathcal{B}}^V = &~\frac{ia_1\mathfrak{B}{\mu_{(F)}}}{a_2\pk B}\left(B - 3\Sigma\left(\Sigma + \frac{\Theta}{3}\right)\right){\mathcal{V}_{(F)}}^S  \\
		~ 
		& -\frac{3\mathfrak{B}\Sigma}{B}{\mu_{(F)}}^V +3a_2\mathfrak{B}\Sigma C\mathcal{E}^T - a_2\mathfrak{B}B C \Sigma^T  - \frac{ia_2\pk\mathfrak{B}J}{a_1}\overline{\mathcal{H}}^T\\
		~ 
		&-\frac{2ia_1\mathfrak{B}G}{\pk B}\overline{\Omega}^V + \left( \frac{3\Sigma\mathfrak{B}^2}{B}\left(\Sigma - \frac{2\Theta}{3}\right) - \frac{\okk}{a_2^2}\right)\overline{\mathfrak{E}}^V \\
~ 
		&+ \frac{3}{2}\left(\Sigma - \frac{2\Theta}{3} - \frac{2\Sigma\mathfrak{B}^2}{B}\right)\mathcal{B}^V - \frac{3\mathfrak{B}\Sigma{\mu_{(F)}}}{B}\left(\Sigma - \frac{2\Theta}{3}\right){\mathcal{V}_{(F)}}^V,
	\end{split} \\
	~
	\begin{split}
		\dot{\overline{\mathcal{H}}}^T = & - \frac{i\pk}{a_1a_2B}\left(1-\frac{2 a_2^2\mathfrak{B}^2}{\okk}\right)\left({\mu_{(F)}}^V + \mathfrak{B}\mathcal{B}^V\right) + \frac{ia_2\pk}{a_1\okk}\mathfrak{B}\mathcal{B}^V \\
		~ 
		&+\frac{2}{a_2B}\left(\mu+p+\Pi\right)\left(\Sigma + \frac{\Theta}{3} \right)\left(1 - \frac{2a_2^2\mathfrak{B}^2}{\okk}\right)\overline{\Omega}^V   \\
		~ 
		&+\frac{1}{a_2^2B}\left( 1-\frac{2a_2^2\mathfrak{B}^2}{\okk}\right)\left(\Sigma + \frac{\Theta}{3}\right) {\mu_{(F)}}{\mathcal{V}_{(F)}}^S \\
		~
		& - \frac{i\pk}{a_1a_2B}\left(1-\frac{2 a_2^2\mathfrak{B}^2}{\okk } \right)\left(\Sigma - \frac{2\Theta}{3}\right)\left({\mu_{(F)}}{\mathcal{V}_{(F)}}^V -  \mathfrak{B}\overline{\mathfrak{E}}^V\right)\\
		~ 
		&+\frac{ia_1}{\pk}\left(\frac{a_2^2\mathfrak{B}^2BL}{3\okk} +\frac{3\Pi}{2}\left(\Sigma + \frac{\Theta}{3}\right)\right)\Sigma^T - \left(\frac{2a_2^2\mathfrak{B}^2L}{3\okk}+\frac{3}{2}\left(F + \frac{M}{B}\right)\right)\overline{\mathcal{H}}^T \\
		~
		& - \frac{ia_1}{\pk}\left(\frac{3\Pi}{2}+  \left(1- C\right)\frac{\pkk}{a_1^2} - \frac{\mathfrak{B}^2}{a_2}\left( \frac{\mathcal{R}a_2^2-\okk}{a_2B} - \frac{a_2^3\Sigma L}{\okk}\right) \right)\mathcal{E}^T,
	\end{split}  \\
			~ 
	\begin{split}
		\dot{\mathcal{V}}\indices{_{(F)}^S} =& - \left(\Sigma + \frac{\Theta}{3}\right){\mathcal{V}_{(F)}}^S,
	\end{split} \label{eq:VFs} \\
	~ 
	\begin{split}
		\dot{\mathcal{V}}\indices{_{(F)}^V} = &~ \frac{1}{2}\left(\Sigma- \frac{2\Theta}{3} - \frac{2\mathfrak{B}^2}{\eta{\mu_{(F)}}}\right){\mathcal{V}_{(F)}}^V - \frac{\mathfrak{B}}{\eta{\mu_{(F)}}}\overline{\mathfrak{E}}^V,
	\end{split} \label{eq:VFV} \\
		~ 
	\begin{split}
		\dot{\mathcal{E}}^T =  &-\frac{3}{2} \left( F + \Sigma C\right)\mathcal{E}^T +\frac{i\pk}{2a_1}\left(J-2\right)\overline{\mathcal{H}}^T \\
		&- \frac{1}{2}\left(\mu + p - 2\Pi\right)\Sigma^T + \frac{3\Sigma}{2a_2B}\left({\mu_{(F)}}^V + \mathfrak{B}\mathcal{B}^V\right) + \frac{ia_1G}{a_2\pk B}\overline{\Omega}^V \\
		&- \frac{ia_1}{2a_2^2\pk B}\left(B - 3\Sigma\left(\Sigma + \frac{\Theta}{3}\right)\right) {\mu_{(F)}}{\mathcal{V}_{(F)}}^S \\
		&- \frac{1}{a_2B}\left(B - \frac{3\Sigma}{2}\left(\Sigma - \frac{2\Theta}{3}\right)\right)\left({\mu_{(F)}}{\mathcal{V}_{(F)}}^V -  \mathfrak{B}\overline{\mathfrak{E}}^V\right),
	\end{split}   \\
			~ 
	\begin{split}
		\dot{\overline{\mathfrak{E}}}^V = &~\frac{2a_2^2\pkk\mathfrak{B}}{a_1^2\okk B}{\mu_{(F)}}^V + \frac{2ia_2\pk\mathfrak{B}{\mu_{(F)}}}{a_1\okk B}\left(\Sigma + \frac{\Theta}{3}\right){\mathcal{V}_{(F)}}^S \\
		~ 
		& + \left( \frac{2a_2^2\pkk{\mu_{(F)}}}{a_1^2\okk B}\left(\Sigma - \frac{2\Theta}{3}\right) - \frac{1}{\eta}\right)\mathfrak{B}{\mathcal{V}_{(F)}}^V  \\
		~ 
		& + \left( \frac{\mathcal{R}a_2^2-\okk}{a_2 B} - \frac{a_2^3\Sigma L}{\okk}\right)\mathfrak{B}\mathcal{E}^T + \frac{a_2^3\mathfrak{B}BL}{3\okk}\Sigma^T + \frac{2ia_2^3\pk\mathfrak{B}L}{3a_1 \okk}\overline{\mathcal{H}}^T \\
		~ 
		& + \frac{4ia_2^2\pk\mathfrak{B}}{a_1\okk B}\left(\mu+p+\Pi\right)\left(\Sigma+\frac{\Theta}{3}\right)\overline{\Omega}^V  + \left( 1+ \frac{a_2^2\pkk}{a_1^2\okk}\left(1+\frac{2\mathfrak{B}^2}{B}\right)\right)\mathcal{B}^V\\
		~
		&-\left(\frac{\Sigma}{2} + \frac{2\Theta}{3} + \frac{1}{\eta} + \frac{2a_2^2\pkk\mathfrak{B}^2}{a_1^2\okk B}\left(\Sigma - \frac{2\Theta}{3}\right)\right)\overline{\mathfrak{E}}^V,
	\end{split} \label{ps:eq23}  \\
	~
	\dot{\Sigma}^T = &~ \left(\Sigma - \frac{2\Theta}{3}\right)\Sigma^T - \mathcal{E}^T, \\
	~
	\dot{\overline{\Omega}}^V =&  - \left( \frac{2\Theta}{3} + \frac{\Sigma}{2}\right)\overline{\Omega}^V, \label{eq:EvenODELast}
\end{align}
where the additional auxiliary variables are defined as
\begin{align}
	M =&~ 2\left(\mathcal{E}+ \frac{\Pi}{2}\right)\left(\Sigma + \frac{\Theta}{3}\right)  +\frac{\Sigma}{a_2^2}\left(\mathcal{R}a_2^2 - \okk \right), \\
	~
	JB =&~ \frac{a_1^2\okk }{a_2^2\pkk}\left(\mathcal{R} - \tilde{k}^2\right)  - 3\Sigma\left(\Sigma - \frac{2\Theta}{3}\right), \\
	~ 
	G =&~ \left(\mu + p + \Pi\right)\left(\mathcal{R} - \tilde{k}^2\right), \\
	~ 
	LB =&~ 3\Sigma\left(\frac{\okk}{a_2^2} - \frac{\pkk}{a_1^2}\right) + \Theta\tilde{k}^2. 
\end{align}

The remaining relavant coefficients can be written in the following way in terms of $\overline{\Omega}^V$, $\overline{\mathcal{H}}^T$, $\mathcal{E}^T$, $\Sigma^T$, ${\mu_{(F)}}^V$, ${\mathcal{V}_{(F)}}^S$, ${\mathcal{V}_{(F)}}^V$, $\mathcal{B}^V$, and $\overline{\mathfrak{E}}^V$. The scalar coefficients are
\begin{align}
Q^S =&~ \mu_{(F)}\mathcal{V}\indices{_{(F)}^S}, \\
\begin{split}
	\frac{i\pk}{a_1a_2^2}\phi^S  = &- \frac{BL}{3}\Sigma^T - \frac{2\pkk}{a_1^2a_2B}\left(\mu\indices{_{(F)}^V} + \mathfrak{B}\mathcal{B}^V\right)  - \frac{2i\pk L}{3a_1}\overline{\mathcal{H}}^T \\
	& -\frac{2i\pk}{a_1a_2}\left( \Sigma-\frac{2\Theta}{3} + \frac{2\left(\mu + p + \Pi\right)}{B}\left(\Sigma+\frac{\Theta}{3}\right)\right)\overline{\Omega}^V  \\
	&+ \left( \Sigma L - \frac{\okk}{a_2^4B}\left(\mathcal{R}a_2^2 - \okk \right) \right)\mathcal{E}^T - \frac{2i\pk}{a_1a_2^2B}\left(\Sigma + \frac{\Theta}{3}\right) \mu_{(F)}\mathcal{V}\indices{_{(F)}^S} \\
	&- \frac{2\pkk}{a_1^2a_2B}\left(\Sigma - \frac{2\Theta}{3}\right)\left(\mu_{(F)}\mathcal{V}\indices{_{(F)}^V} -  \mathfrak{B}\overline{\mathfrak{E}}^V\right).
\end{split} 
\end{align}
The vector coefficients are in turn given by
\begin{align}
\mu^V =&~ \mu\indices{_{(F)}^V} + \mathfrak{B}\mathcal{B}^V, \\
~
p^V =&~\frac{\mathfrak{B}}{3}\mathcal{B}^V, \quad Y^V = -\frac{4\mathfrak{B}}{3}\mathcal{B}^V, \\
~
Q^V =&~ \mu_{(F)}\mathcal{V}\indices{_{(F)}^V} -  \mathfrak{B}\overline{\mathfrak{E}}^V, \\
~
\begin{split}
\mathfrak{B}^V = &~\frac{2i\pk a_2^2\mathfrak{B}}{a_1\okk B}\left(\mu\indices{_{(F)}^V}+\mathfrak{B}\mathcal{B}^V\right) + \frac{i\pk a_2^2}{a_1\okk}\mathcal{B}^V  \\
~ 
& - \frac{4a_2^2\mathfrak{B}}{\okk B}\left(\mu+p+\Pi\right)\left(\Sigma+\frac{\Theta}{3}\right) \overline{\Omega}^V\\
~ 
& -\frac{2a_2^3\mathfrak{B}L}{3\okk}\overline{\mathcal{H}}^T + \frac{i a_1a_2^3\mathfrak{B}BL}{3\pk\okk}\Sigma^T  \\
~ 
&+\frac{ia_1\mathfrak{B}}{\pk}\left( \frac{\mathcal{R}a_2^2-\okk}{a_2B} - \frac{a_2^3\Sigma L}{\okk}\right)\mathcal{E}^T -\frac{2a_2\mathfrak{B}}{\okk B}\left(\Sigma + \frac{\Theta}{3}\right)\mu_{(F)}\mathcal{V}\indices{_{(F)}^S} \\
~ 
& +\frac{2i\pk a_2^2\mathfrak{B}}{\okk a_1 B}\left(\Sigma - \frac{2\Theta}{3}\right)\left(\mu_{(F)}\mathcal{V}\indices{_{(F)}^V}-\mathfrak{B}\overline{\mathfrak{E}}^V\right),
\label{eq:Bv}
\end{split}\\ 
~ 
\Pi^V =& -\mathfrak{B}\mathfrak{B}^V,\\ 
~ 
\begin{split}
\frac{V^V}{a_2}	= &-\frac{B}{3}\left(2C+1\right)\Sigma^T - \frac{2i\pk}{3a_1}\left(1+J\right)\overline{\mathcal{H}}^T + \Sigma\left(2C+1\right)\mathcal{E}^T \\
 	&-\frac{2\Sigma}{a_2B}\left(\mu\indices{_{(F)}^V} + \mathfrak{B}\mathcal{B}^V\right) + \frac{4ia_1}{3a_2\pk}\left( \frac{\pkk}{a_1^2} - \frac{G}{B}\right)\overline{\Omega}^V \\
 	&+ \frac{2ia_1}{3a_2^2\pk B}\left(B- 3\Sigma\left(\Sigma + \frac{\Theta}{3}\right) \right) \mu_{(F)}\mathcal{V}\indices{_{(F)}^S} \\
 	&+ \frac{2}{3a_2B}\left(B - 3\Sigma\left(\Sigma - \frac{2\Theta}{3}\right) \right)\left(\mu_{(F)}\mathcal{V}\indices{_{(F)}^V} -  \mathfrak{B}\overline{\mathfrak{E}}^V\right),
\end{split} \\
~
\begin{split}
	\frac{W^V}{a_2}= &\frac{B}{2}\left(C-1\right)\Sigma^T + \frac{3\Sigma}{2a_2B}\left(\mu\indices{_{(F)}^V} + \mathfrak{B}\mathcal{B}^V\right) + \frac{i\pk}{2a_1}\left(J-2\right)\overline{\mathcal{H}}^T \\
	&+ \frac{ia_1}{a_2\pk}\left(\frac{2\pkk}{a_1^2} + \frac{G}{B}\right)\overline{\Omega}^V + \frac{3\Sigma}{2}\left(1-C\right)\mathcal{E}^T  \\
	&- \frac{ia_1}{2a_2^2\pk B}\left( B - 3\Sigma\left(\Sigma + \frac{\Theta}{3}\right)\right) \mu_{(F)}\mathcal{V}\indices{_{(F)}^S}   \\
	&+\frac{1}{a_2B}\left(B + \frac{3\Sigma}{2}\left(\Sigma - \frac{2\Theta}{3}\right) \right) \left(\mu_{(F)}\mathcal{V}\indices{_{(F)}^V} -  \mathfrak{B}\overline{\mathfrak{E}}^V\right),
\end{split} \\
~
\begin{split}
	\frac{i\pk}{a_1a_2}\mathcal{E}^V = &\frac{\pkk}{a_1^2a_2B}\left(1-\frac{ a_2^2\mathfrak{B}^2}{\okk}\right)\left(\mu\indices{_{(F)}^V} + \mathfrak{B}\mathcal{B}^V\right)-\frac{a_2\pkk }{2a_1^2\okk}\mathfrak{B}\mathcal{B}^V \\
	~~
 &-\left( \frac{ a_2^2\mathfrak{B}^2BL}{6\okk}+ \frac{3}{2}\left(\mathcal{E} + \frac{\Pi}{2} \right)\left(\Sigma + \frac{\Theta}{3} \right)\right)\Sigma^T \\
	~~
&+\frac{2i\pk}{a_1 a_2B} \left(\mu +p +\Pi\right)\left(\Sigma + \frac{\Theta}{3} \right) \left(1  -\frac{a_2^2\mathfrak{B}^2}{\okk} \right)\overline{\Omega}^V  \\
     ~~
	& -\frac{i\pk}{a_1}\left( \frac{3}{2B}\left( \frac{\Sigma}{a_2^2}\left(\mathcal{R}a_2^2 - \okk\right) + 2\left(\mathcal{E} + \frac{\Pi}{2}\right)\left(\Sigma + \frac{\Theta}{3}\right) \right)+\frac{a_2^2\mathfrak{B}^2L}{3\okk}\right)\overline{\mathcal{H}}^T \\
	~~
		& + \left( \frac{3}{2}\left(\mathcal{E} + \frac{\Pi}{2}\right) - \frac{\pkk}{a_1^2}C - \frac{\mathfrak{B}^2}{2} \left( \frac{\mathcal{R}a_2^2-\okk}{a_2^2B} - \frac{a_2^2\Sigma L}{\okk}\right) \right)\mathcal{E}^T \\
		~
		&+ \frac{i\pk}{a_1a_2^2B} \left(1-\frac{a_2^2\mathfrak{B}^2}{\okk } \right)\left(\Sigma + \frac{\Theta}{3}\right) \mu_{(F)}\mathcal{V}\indices{_{(F)}^S} \\
		~ 
		& + \frac{\pkk}{a_1^2a_2B} \left(1- \frac{a_2^2\mathfrak{B}^2}{\okk} \right)\left(\Sigma - \frac{2\Theta}{3} \right)\left(\mu_{(F)}\mathcal{V}\indices{_{(F)}^V} -  \mathfrak{B}\overline{\mathfrak{E}}^V\right),
\end{split} \\
~
\begin{split}
	\frac{X^V}{a_2}= &\frac{3}{2}\left(\Sigma - \frac{2\Theta}{3}\right)\left(\mathcal{E} + \frac{\Pi}{2}\right)\Sigma^T + \frac{1}{a_2}\left( 1 - \frac{1}{B}\left( \frac{\okk}{a_2^2} + 3\left(\mathcal{E} + \frac{\Pi}{2} \right) \right) \right)\mathfrak{B}\mathcal{B}^V\\
	~
	&+ \frac{1}{3a_2}\left( 1 - \frac{3}{B}\left( \frac{\okk}{a_2^2} + 3\left(\mathcal{E} + \frac{\Pi}{2} \right) \right) \right)\mu\indices{_{(F)}^V}\\
	& +\frac{3ia_1}{\pk B}\left( \frac{\pkk}{a_1^2}\left(\mathcal{E} + \frac{\Pi}{2} \right)\left(\Sigma - \frac{2\Theta}{3} \right) + \frac{\okk \Sigma}{2a_2^4}\left(\mathcal{R}a_2^2 - \okk\right) \right)\mathcal{\overline{H}}^T \\
	&+ \frac{4i\pk}{a_1a_2B}\left( \left(\mu + p + \Pi\right)\left(\Sigma + \frac{\Theta}{3}\right) + \frac{3a_1^2\Sigma G}{4\pkk} \right) \overline{\Omega}^V \\
	&+ C\left( \frac{\okk}{a_2^2} + 3\left(\mathcal{E} + \frac{\Pi}{2} \right) \right)\mathcal{E}^T  \\
	& + \frac{ia_1}{a_2^2\pk B}\left( \frac{3\Sigma}{2a_2^2}\left(\mathcal{R}a_2^2 - \okk \right) - \frac{\pkk}{a_1^2}\left(\Sigma - \frac{2\Theta}{3}\right) \right) \mu_{(F)}\mathcal{V}\indices{_{(F)}^S}  \\
	& + \frac{1}{a_2B}\left(\Sigma - \frac{2\Theta}{3} \right)\left( \mathcal{R} - \frac{\okk}{a_2^2} +\frac{3\Sigma}{2}\left(\Sigma - \frac{2\Theta}{3}\right) \right) \left(\mu_{(F)}\mathcal{V}\indices{_{(F)}^V} -  \mathfrak{B}\overline{\mathfrak{E}}^V\right),
\end{split} \\
~ 
\begin{split}
		\frac{i\pk}{a_1a_2}\Sigma^V = &~\frac{i\pk}{a_1a_2}\overline{\Omega}^V - \frac{1}{2}\left(\mathcal{R} - \frac{\okk}{a_2^2} + B\right)\Sigma^T \\
		&+ \frac{3\Sigma}{2}\mathcal{E}^T- \frac{i\pk}{a_1}\overline{\mathcal{H}}^T,
\end{split}\\ 
~ 
\begin{split}
\frac{2}{3a_2}\overline{\mathcal{H}}^V = &-\left(\mathcal{E} + \frac{\Pi}{2}\right)\Sigma^T - \frac{\Sigma}{a_2B}\left(\mu\indices{_{(F)}^V} + \mathfrak{B}\mathcal{B}^V\right) - \frac{i\pk J}{3a_1}\overline{\mathcal{H}}^T - \frac{2ia_1G}{3a_2\pk B}\overline{\Omega}^V \\
	&+ \Sigma C\mathcal{E}^T - \frac{ia_1}{3a_2^2\pk B}\left( 3\Sigma\left(\Sigma + \frac{\Theta}{3}\right) - B \right) \mu_{(F)}\mathcal{V}\indices{_{(F)}^S}  \\
	&- \frac{1}{3a_2B}\left(3\Sigma\left(\Sigma - \frac{2\Theta}{3}\right)- B \right)\left(\mu_{(F)}\mathcal{V}\indices{_{(F)}^V} -  \mathfrak{B}\overline{\mathfrak{E}}^V\right),
\end{split}\\ 
~ 
\begin{split}
	\frac{i\pk}{a_1a_2}\alpha^V = &- \frac{3}{2}\left( \mathcal{E} + \frac{\Pi}{2}\right)\Sigma^T - \frac{3\Sigma}{2a_2B}\left(\mu\indices{_{(F)}^V} + \mathfrak{B}\mathcal{B}^V\right) - \frac{i\pk}{2a_1}J\overline{\mathcal{H}}^T - \frac{ia_1G}{a_2\pk B}\overline{\Omega}^V \\
	& + \frac{3\Sigma C}{2}\mathcal{E}^T + \frac{ia_1}{2a_2^2\pk B}\left( B - 3\Sigma\left(\Sigma + \frac{\Theta}{3}\right)\right) \mu_{(F)}\mathcal{V}\indices{_{(F)}^S}  \\
	&+ \frac{1}{a_2B}\left(B- \frac{3\Sigma}{2}\left(\Sigma - \frac{2\Theta}{3}\right)\right)\left(\mu_{(F)}\mathcal{V}\indices{_{(F)}^V} -  \mathfrak{B}\overline{\mathfrak{E}}^V\right).
	\end{split}
\end{align}
Finally, the only remaining tensorial coefficient is
\begin{equation}
\frac{i\pk}{a_1}\zeta^T  =~ \left(\Sigma + \frac{\Theta}{3}\right)\Sigma^T - \mathcal{E}^T.
\end{equation}

\section{Analyzing the Final System}\label{sec:Analyzing the Final System}

Since the final equations from the previous section are still rather difficult to deal with analytically, we will mainly consider numerical examples. In doing so, we will focus on analyzing  how the perturbations depend on the chosen length scales. However, before moving on to the numerical calculations, we should make some notes about the inherent wave length dependencies introduced when defining the gauge invariant variables and the harmonics.

\subsection{Rescaling the Harmonics}
When investigating the dependence on length scale, it is important to note that, due to our definition of the vector and tensor harmonics as gradients of the scalar harmonics, these will naturally contain factors of $\ok$. Hence, the vector and tensor coefficients will contain inherent factors of $\ok$ relative to the scalar coefficients. Defining a new set of harmonics,  where the inherent factors of $\ok$ have been removed,
\begin{alignat}{2}
Q\indices{_*^{k_{\perp}}_a} &= \frac{a_2}{\ok}\tensor{\delta}{_a}Q^{k_{\perp}}, \quad &Q\indices{_*^{k_{\perp}}_a_b} &= \frac{a_2^2}{\okk}\tensor{\delta}{_{\{a}}\tensor{\delta}{_{b\}}}Q^{k_{\perp}},\\
\overline{Q}\indices{_*^{k_{\perp}}_a}  &= \frac{a_2}{\ok}\tensor{\epsilon}{_a_b}\tensor{\delta}{^b}Q^{k_{\perp}},  \quad & \overline{Q}\indices{_*^{k_{\perp}}_a_b} &= \frac{a_2^2}{\okk}\tensor{\epsilon}{_c_{\{a}}\tensor{\delta}{^c}\tensor{\delta}{_{b\}}}Q^{k_{\perp}}.
\end{alignat}
we note that 
\begin{alignat}{2}
\Psi\indices{^V} &=k^{-1}_{\perp}\Psi\indices{_*^V} , \quad & \overline{\Psi}\indices{^V} &= k^{-1}_{\perp} \overline{\Psi}\indices{_*^V} ,\\
\Psi\indices{^T} &=k^{-2}_{\perp}\Psi\indices{_*^T} , \quad & \overline{\Psi}\indices{^T} &= k^{-2}_{\perp} \overline{\Psi}\indices{_*^T},
\end{alignat}
where the $\Psi_*$ are the coefficients relative to the new harmonics.  These harmonics have the benefit that the scalar, vector and tensor coefficients do not differ with factors of $\ok$ simply due to the definition of the harmonics. 

However, there are still some inherent factors of $\ok$ that need to be addressed. These factors occur for the gauge invariant variables that are defined as gradients of quantities that are non-zero on the background spacetime.  If we let $\Phi_a = \delta_a \Phi$, then the $\delta$ gradient implies that the coefficients $\{ \Phi\indices{_*^V}, \overline{\Phi}\indices{_*^V} \}$ will naturally be one order higher in the factor $\ok/a_2$ than the coefficients for the corresponding scalar perturbation of $\Phi$. Hence, when investigating the length scale dependencies, we will focus on coefficients of the form $\{ \Psi^S, \Psi\indices{_*^{V,T}}, \overline{\Psi}\indices{_*^{V,T}},  a_2k^{-1}_{\perp}\Phi\indices{_*^V}, a_2k^{-1}_{\perp}\overline{\Phi}\indices{_*^V}\}$ to avoid the aforementioned inherent factors of $\ok$ due to our definitions.

\subsection{Dependence on Perturbation Length Scale}\label{sec:Dependence on Perturbation Length Scale}
Equipped with the full system of ordinary differential equations and a rescaled set of harmonic coefficients, we now consider specific numerical examples. In these calculations, we will follow a similar convention as in \cite{Bradley_2011} and set $\Lambda = 1$, which will fixate the remaining length dimension. As such, we will in the following treat all quantities as being dimensionless. In our calculations we will use the background spacetime specified by the following initial conditions and parameter values
\begin{alignat}{5}
\Sigma(t_0) &= 0, \quad  & \mu_{(F)}(t_0)&=(1-r_\mathfrak{B})\mu(t_0),   \quad  & a_1(t_0) &= 1, \notag \\
 \Theta(t_0)&=0.6, \quad  &\mathfrak{B}(t_0)&=\sqrt{2r_\mathfrak{B}\mu(t_0)}, \quad  & a_2(t_0) &=1, \label{eq:InitialBackground} \\
\mu(t_0)&=0.36,  \quad  &  \gamma&= \frac{5}{3},   \quad  & &\notag 
\end{alignat}
where $r_\mathfrak{B}$ is the fraction between the initial magnetic energy density and the total energy density. The initial conditions for $\mu$, $\Sigma$, and $\Theta$ are identical to those for a dust background presented in \cite{Bradley_2011}, which ends at an expanding de Sitter solution. When not stated otherwise, we will use $r_\mathfrak{B} = 0.1$ and the initial resistivity $\eta(t_0)=10^{-3}$. We will also choose the $k$-space direction $\pk = 2\ok = 2k(t_0)/\sqrt{5}$, where 
\begin{equation}
k^2 = \frac{\pkk}{a_1^2} + \frac{\okk}{a_2^2},
\end{equation}
is the square of the norm of the physical wave vector. For all numerical calculations, we solve Eqs.~(\ref{eq:Theta_dot}), (\ref{eq:Sigma_dot}), (\ref{eq:muF_dot}), (\ref{eq:b_dot}), (\ref{eq:OddODEFirst})–(\ref{eq:OddODELast}), and (\ref{eq:EvenODEFirst})–(\ref{eq:EvenODELast}) together with Eqs.~(\ref{eq:adot}), and (\ref{eq:etadot}). When not stated otherwise, the equations are solved using \texttt{ode45} in \texttt{MATLAB} with a relative tolerance of $10^{-9}$ and an absolute tolerance of $10^{-34}$.

\subsubsection{Tensor Perturbations}
\begin{figure}[t]
\centering
\includegraphics[scale=1]{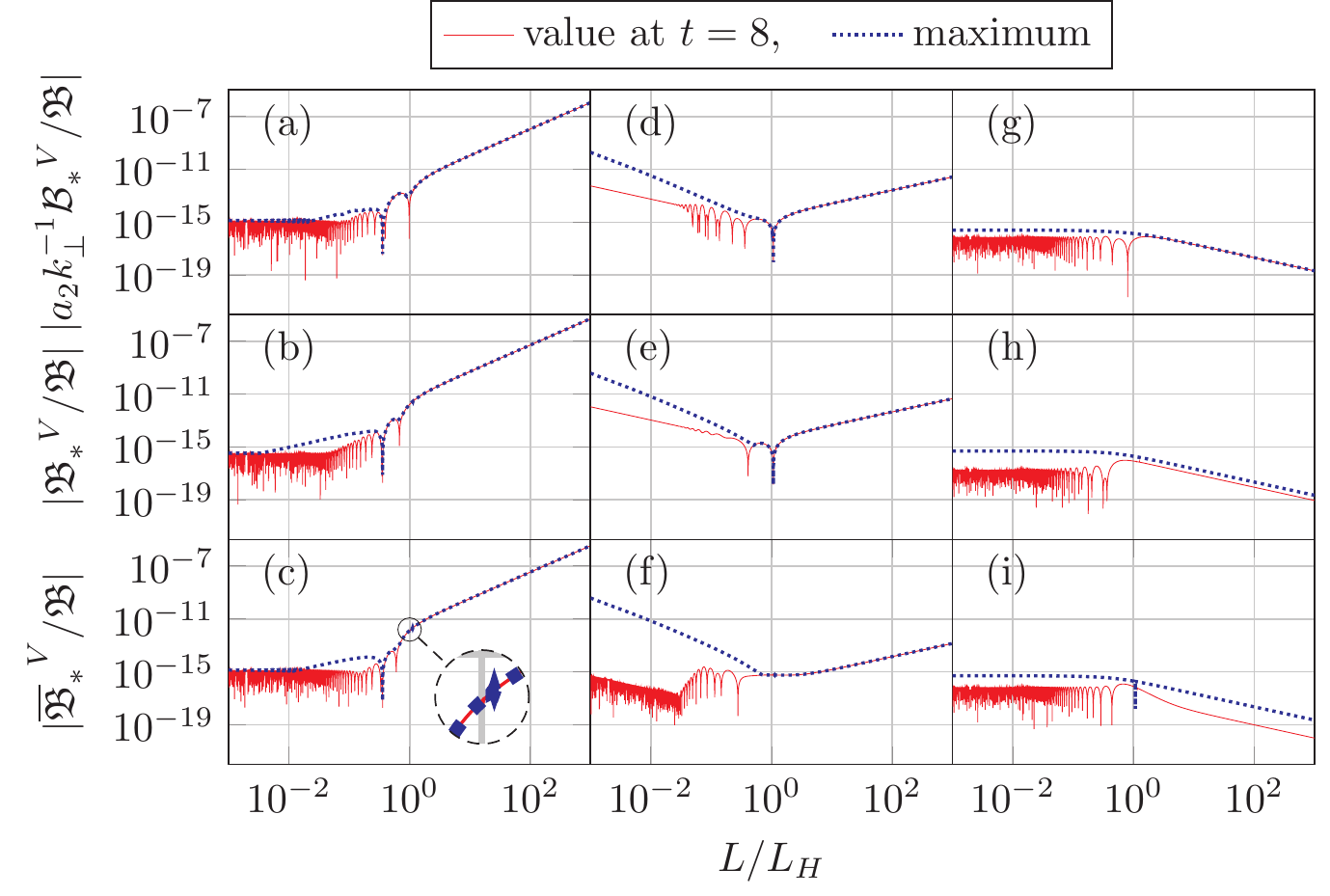}
\caption{Generated magnetic fields as functions of perturbation length scale using the background spacetime specified in Eq.~(\ref{eq:InitialBackground}). Both the maximum amplitude and the values extracted at the integration time $t=8$ are shown. In a–c the results using the initial conditions specified in Eqs.~(\ref{eq:InitialPert1start})–(\ref{eq:InitialPert1end}) are displayed, whilst d–f and g–i uses  Eqs.~(\ref{eq:InitialPert2start})–(\ref{eq:InitialPert2end}) and Eqs.~(\ref{eq:InitialPert3start})–(\ref{eq:InitialPert3end})  respectively. In all calculations, $\pk = 2\ok = 2k(t_0)/\sqrt{5}$, and $r_{\mathfrak{B}} = 0.1$. The initial resistivity was $\eta(t_0)=10^{-3}$ in a–f and $\eta(t_0)=1$ in g–h. Sharp and noise-like spikes slightly above $L/L_H$ equal to unity may be due to numerical problems encountered when the auxiliary function $B$ approaches zero. }
\label{fig:1}
\end{figure}

As a first example, we consider the generated magnetic fields due to perturbations in some tensorial coefficients. Following a similar line as in \cite{Zunckel_2006} we will let the initial tensorial shear perturbation represent some initial gravitational wave content. The aim is then to investigate how the generated magnetic fields depend on the chosen length scale. Defining the initial Hubble length $L_H  = 3/\Theta(t_0)$ and a characteristic length scale for the perturbations $L = 2\pi/k(t_0)$, we solve the final system of equations for values of $L/L_H$ in the range $\left[10^{-3}, 10^{3} \right]$. As for the initial perturbations, we choose  
\begin{align}
\Sigma\indices{_*^T}(t_0) = \overline{\Sigma}\indices{_*^T}(t_0) =  10^{-19}k_m^2 \sim 10^{-13}, 
\label{eq:InitialPert1start}
\end{align}
or equivalently
\begin{align}
\Sigma\indices{^T}(t_0) = \overline{\Sigma}\indices{^T}(t_0) =  10^{-19}\frac{k_m^2}{\okk}, 
\end{align}
where $k_m$ is the maximum value of $k(t_0)$ in the studied range. Requiring $\mathfrak{B}^V(t_0) = 0$ and using Eqs.~(\ref{eq:barSigmaT}) and (\ref{eq:Bv}), we also let 
\begin{equation}
\mathcal{H}^T(t_0) = -\frac{ia_1B}{2\pk}\overline{\Sigma}^T\Bigg\rvert_{t_0}, \quad \overline{\mathcal{H}}^T(t_0) = \frac{ia_1B}{2\pk}\Sigma^T\Bigg\rvert_{t_0},
\label{eq:InitialPert1end}
\end{equation} 
whilst $\{\Omega^S$, $\overline{\mathcal{E}}^T$, ${\overline{\mathcal{V}}_{(F)}}^V$, $\mathfrak{E}^S$, $\mathfrak{E}^V$, $\overline{\mathfrak{B}}^V$, $\overline{\Omega}^V$, $\mathcal{E}^T$, ${\mu_{(F)}}^V$, ${\mathcal{V}_{(F)}}^S$, ${\mathcal{V}_{(F)}}^V$, $\mathcal{B}^V$, $\overline{\mathfrak{E}}^V \}$ are chosen to vanish initially. Hence, in particular, the perturbations of the electromagnetic fields and the plasma velocities are all zero at $t_0$. Solving the system, we then get the results shown in Fig.~\ref{fig:1}a–c.  For super-horizon scales,  $L \gg L_H$, the curves in the logarithmic plot are linear with a slope close to two. Decreasing $L$ below $L_H$, the values extracted at the fixed time $t = 8$ begin to oscillate. In the region around $L/L_H = 10^{-1}$, the curves displaying the maximum values are almost linear with a slope of approximately one. Eventually when decreasing $L$ even further, it can be seen that the curves approach a constant value. 
Finally, it should be remarked that the noise-like spikes at values of $L/L_H$ slightly larger than unity, as can be seen in Fig.~\ref{fig:1}c, are most likely due to numerical problems encountered when the auxiliary variable $B$, defined in Eq.~(\ref{eq:auxB}), passes through zero for $t>t_0$. As the final system of ordinary differential equations have been derived under the assumption that $B$ is non-zero, a zero-crossing for $B$ clearly poses a problem, as the derivatives involving factors of $B^{-1}$ are not well defined when $B=0$. In addition to the derivatives not being well defined, we also encounter problems with the coefficients that are given as algebraic expressions in terms of the coefficients in the final system, as their corresponding expressions may involve factors of $B^{-1}$. Therefore, we also have to be careful when specifying initial conditions for the perturbations when $B(t_0) \sim 0$, since the coefficients with algebraic expressions, which should remain small relative to the background, may become too large even if the initial conditions for the variables in the ODE system appear small.

\subsubsection{Velocity Perturbations With Vanishing Initial Ohmic Current}\label{sec:Velocity Perturbations With Vanishing Initial Ohmic Current}
As a second example, we now consider how the generated magnetic field is affected by perturbations in the plasma velocity. To avoid aforementioned problems with the initial conditions when $B(t_0) \sim 0$, we write the initial velocity perturbation as \footnote{ This can be motivated by observing equations such as Eq.~(\ref{eq:barSigmaT}). In this specific scenario, adding the extra factor $B(t_0)$ seems to be most important for the magnetic fields in the odd sector, as performing the calculations without this factor causes a sharp increase in $\overline{\mathfrak{B}}\indices{_*^V}/\mathfrak{B}$ near $L/L_H$ equal to unity. However, no such obvious increase is observed for the even magnetic fields other than the previously mentioned noise-like spikes.} 
\begin{equation}
\mathcal{V}\indices{_{(F)}_*^V}(t_0) = \overline{\mathcal{V}}\indices{_{(F)}_*^V}(t_0) = 10^{-19}k_mB(t_0).
\label{eq:InitialPert2start}
\end{equation}
Since the electrical resistivity is rather small, we also introduce the following initial perturbations in the electric field 
\begin{equation}
\mathfrak{E}^V(t_0) = \mathfrak{B}\overline{\mathcal{V}}\indices{_{(F)}^V}\big\rvert_{t_0} , \quad \overline{\mathfrak{E}}^V(t_0) = -\mathfrak{B}\mathcal{V}\indices{_{(F)}^V}\big\rvert_{t_0}
\end{equation}
as to avoid large initial ohmic currents. Finally, requiring that $\mathfrak{B}^V(t_0) = 0$, we use Eq.~(\ref{eq:Bv}) and choose 
\begin{equation}
\mathcal{V}\indices{_{(F)}^S}(t_0) = \frac{i\pk a_2}{a_1}\left(\Sigma-\frac{2\Theta}{3}\right)\left(\Sigma+\frac{\Theta}{3}\right)^{-1}\left( \mathcal{V}\indices{_{(F)}^V} - \frac{\mathfrak{B}}{\mu_{(F)}}\overline{\mathfrak{E}}^V\right) \Bigg \rvert_{t_0}. 
\label{eq:InitialPert2end}
\end{equation}
The other perturbations, $\{\Omega^S$, $\overline{\mathcal{E}}^T$, $\mathcal{H}^T$, $\mathfrak{E}^S$, $\overline{\mathfrak{B}}^V$, $\overline{\Omega}^V$, $\overline{\mathcal{H}}^T$, $\mathcal{E}^T$, $\Sigma^T$,$ {\mu_{(F)}}^V$, $\mathcal{B}^V \}$, that are explicitly present in the ODE system are set to zero initially. Performing the calculations over the same range of length scales as in the previous example, the results in Fig.~\ref{fig:1}d–f are obtained. There it can be seen that for superhorizon scales, the maximum and end values at $t=8$  follow each other closely with a slope of approximately unity in the logarithmic scale. However, for subhorizon scales, the curves displaying the curves seemingly diverge. The maximum curve continues with a slope of roughly $-1.7$ whilst the curve representing the values at $t=8$ follow a slope closer to $-1$. Finally, it should be noted that the values at $t=8$ begin oscillating when decreasing the length scale below the Hubble scale. However, in contrast to the behavior in the odd sector shown in Fig.~\ref{fig:1}f, the oscillations of the quantities in the even sector, displayed in  Fig.~\ref{fig:1}d–e, have become unnoticeable for length scales below $10^{-2}L_H$.

\begin{figure}[h]
\centering
\includegraphics[scale=1]{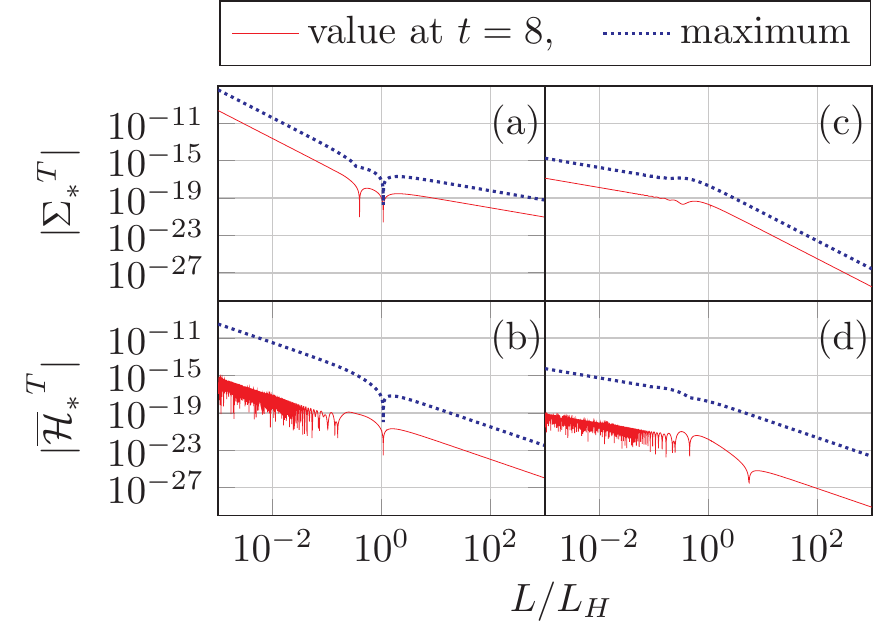}
\caption{Generated tensorial perturbations as functions of the length scale using the background spacetime specified in Eq.~(\ref{eq:InitialBackground}). Both the maximum amplitude and the values extracted at the integration time $t=8$ are shown. In a–b the results using the initial conditions specified in Eqs.~(\ref{eq:InitialPert2start})–(\ref{eq:InitialPert2end}) are displayed, whilst c–d uses Eqs.~(\ref{eq:InitialPert3start})–(\ref{eq:InitialPert3end})  respectively. In all calculations, $\pk = 2\ok = 2k(t_0)/\sqrt{5}$, $r_{\mathfrak{B}} = 0.1$. The initial resistivity was $\eta(t_0)=10^{-3}$ in a–b and $\eta(t_0)=1$ in c–d.}
\label{fig:2}
\end{figure}

Finally, as we are considering initial velocity perturbations, it is also interesting to investigate the generation of perturbations in the tensorial coefficients. In Fig.~\ref{fig:2}a–b we display the generated perturbations in the coefficients $\Sigma\indices{_*^T}$ and $\overline{\mathcal{H}}\indices{_*^T}$. There it can be seen that the generated fields increase with decreasing length scale. For subhorizon scales, both $\Sigma\indices{_*^T}$ and $\overline{\mathcal{H}}\indices{_*^T}$ display straight lines in the logarithmic scale with a slope of approximately $-3$ and $-2$ respectively. On superhorizon scales, the curves for $\Sigma\indices{_*^T}$ have a slope of $-1$ while the curves for $\overline{\mathcal{H}}\indices{_*^T}$ have a slope of about $-2$.

\subsubsection{Velocity Perturbations With Vanishing Initial Energy Flux}
In the previous example, both velocity and electric field perturbations were introduced to ensure that the ohmic currents and magnetic fields vanished initially. However, in doing so we introduce an initial energy flux that
will affect many gravitationally related variables. For instance, this flux appears in both the differential equations and algebraic relations pertaining to the electric and magnetic parts of the Weyl tensor. Hence, the previously chosen initial values may implicitly contain gravitational effects. Therefore, to better isolate the plasma related effects, we also investigate the case where the initial conditions are chosen so that the energy flux vanishes initially. For this purpose we introduce 
\begin{align}
\mathcal{V}\indices{_{(F)}_*^V}(t_0) &= \overline{\mathcal{V}}\indices{_{(F)}_*^V}(t_0) = 10^{-19}k_m \label{eq:InitialPert3start} \\
\mathfrak{E}^V(t_0) &= - \frac{\mu_{(F)}}{\mathfrak{B}}\overline{\mathcal{V}}\indices{_{(F)}^V}\bigg\rvert_{t_0}, \\
\overline{\mathfrak{E}}^V(t_0) &= \frac{\mu_{(F)}}{\mathfrak{B}}\mathcal{V}\indices{_{(F)}^V}\bigg\rvert_{t_0}. \label{eq:InitialPert3end}
\end{align}
where we seemingly no longer need the factor $B(t_0)$ in the velocity perturbations. It should be noted that these initial conditions for the electric field coefficients are incompatible with the ideal MHD limit, as they would in general imply an infinite initial ohmic current for $\eta \rightarrow 0$. To avoid similar large initial ohmic currents in our calculations, we increase the resistivity to $\eta(t_0)=1$. Finally, we set $
\{\Omega^S$, $\overline{\mathcal{E}}^T$, $\mathcal{H}^T$, $\mathfrak{E}^S$, $\overline{\mathfrak{B}}^V$, $\overline{\Omega}^V$, $\overline{\mathcal{H}}^T$, $\mathcal{E}^T$, $\Sigma^T$, ${\mu_{(F)}}^V$, ${\mathcal{V}_{(F)}}^S$, $\mathcal{B}^V \}$ to zero at $t_0$. Then, on performing the calculations in the same range of length scales as in previous examples, we obtain the results shown in Figs.~\ref{fig:1}g–i. There it can be seen that, on superhorizon scales, the curves for the maximum value and the value at $t = 8$ decay with a slope close to $-1$. For subhorizon scales, the maximum curve and the general trend of the oscillating $t=8$ curve  both appear rather independent of the length scale. In Figs.~\ref{fig:1}h–i the values at $t = 8$ are notably smaller than the maximum values, which is also true for subhorizon scales in Fig.~\ref{fig:1}g. For superhorizon scales, the curves in Fig.~\ref{fig:1}g seemingly coincide.

With these initial conditions, it is also interesting to again examine the effect of the velocity perturbations on the tensorial quantities. In Fig.~\ref{fig:2}c–d we display the generated perturbations in the coefficients $\Sigma\indices{_*^T}$ and $\overline{\mathcal{H}}\indices{_*^T}$.  There it can be see that the generated tensorial perturbations increase with decreasing $L/L_H$, with different slopes on super and subhorizon scales.  On superhorizon scales, the curves for $\Sigma\indices{_*^T}$ have a slope of $-3$ while $\overline{\mathcal{H}}\indices{_*^T}$ have a slope of about $-2$ . On subhorizon scales, the curves for $\Sigma\indices{_*^T}$ and $\overline{\mathcal{H}}\indices{_*^T}$ both have a slope of $-1$. Hence, on attributing generated fields in Fig.~\ref{fig:1}g–i and  Fig.~\ref{fig:2} to plasma related effects, it can be seen that these effects become more important on subhorizon scales.

\subsection{Beat Waves}\label{sec:Beat Waves}

\begin{figure}[h]
\centering
\includegraphics[scale=1]{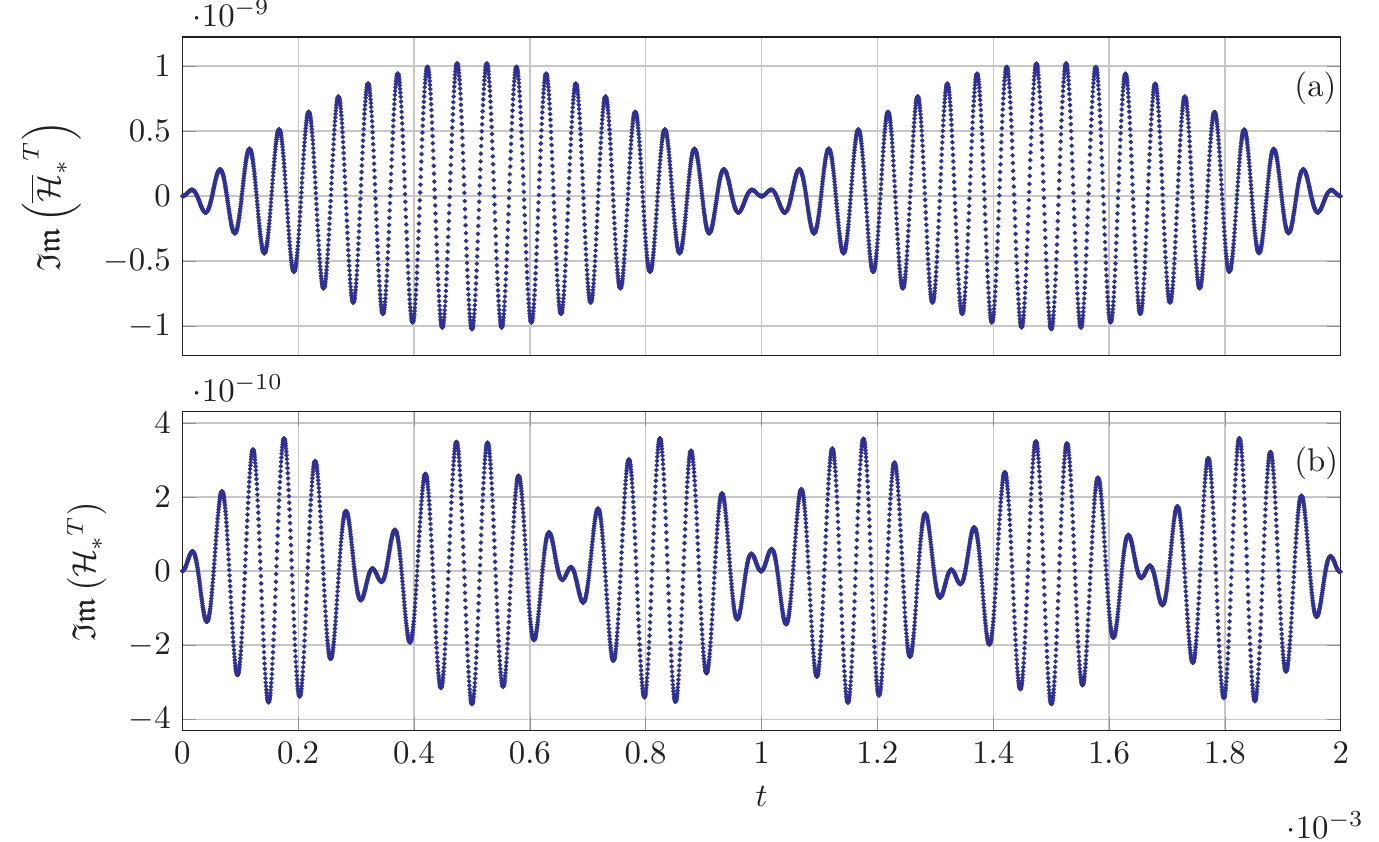}
\caption{Generated tensorial coefficients pertaining to the magnetic part of the Weyl tensor on the perturbation length scale $L = 10^{-5}L_H$ using the background specified in Eq.~(\ref{eq:InitialBackground}). The initial magnitude of the magnetic field on the background was chosen so that $v_A(t_0) = 0.95$. As for the initial perturbations, these were chosen according to Eqs.~(\ref{eq:InitialPert4start})–(\ref{eq:InitialPert4end}). In all calculations, $\pk = 2\ok = 2k(t_0)/\sqrt{5}$ and $\eta(t_0) = 10^{-12}$.} 
\label{fig:3}
\end{figure}

To further investigate the plasma related effects on subhorizon scales, we perform similar calculations as in Sec.~\ref{sec:Velocity Perturbations With Vanishing Initial Ohmic Current}, but varying the magnitude of the background magnetic field rather than the length scale of the perturbations. Hence, we introduce the initial conditions
\begin{align}
\mathcal{V}\indices{_{(F)}_*^V}(t_0) &= \overline{\mathcal{V}}\indices{_{(F)}_*^V}(t_0) = 10^{-10}, \label{eq:InitialPert4start} \\ 
~
\mathfrak{E}^V(t_0) &= \mathfrak{B}\overline{\mathcal{V}}\indices{_{(F)}^V}\big\rvert_{t_0} , \quad \overline{\mathfrak{E}}^V(t_0) = -\mathfrak{B}\mathcal{V}\indices{_{(F)}^V}\big\rvert_{t_0}, \label{eq:InitialPert4end} 
\end{align}
but set $\mathcal{V}\indices{_{(F)}^S}$ to zero. To highlight the sought effects, we decrease the initial resistivity drastically to $\eta(t_0) = 10^{-12}$ and choose the length scale to be $L = 10^{-5}L_H$. Since the smallness of the chosen electrical resistivity appears to make the system quite stiff, we use \texttt{ode15s} instead of \texttt{ode45} in this section, although without specifying an analytical expression for the Jacobian. As for the initial magnitude of the background magnetic field, this is, for reasons made more clear in the next section, specified through the Alfvén velocity $v_A$
\begin{equation}
{v_A}^2 \equiv \cfrac{\cfrac{\mathfrak{B}^2}{ \mu_{(F)}}}{1+\cfrac{\mathfrak{B}^2}{ \mu_{(F)}}}.
\label{eq:Alfven Velocity}
\end{equation}
On using that 
\begin{align}
r_{\mathfrak{B}} = \frac{v_A^2}{2-v_A^2} \Bigg \rvert_{t_0}
\end{align}
in Eq.~(\ref{eq:InitialBackground}) and calculating for $v_A(t_0) = 0.95 $, the results in Fig.~\ref{fig:3} are obtained for the tensorial coefficients of the magnetic part of the Weyl tensor. There it can be seen that a clear beat pattern emerges. 

To get an analytical explanation of this beat pattern, we now seek wave equations for the plasma velocities and the magnetic part of the Weyl tensor in the limit of large wave numbers. Looking at Eq.~\eqref{eq:VFs}, we see that $\mathcal{V}\indices{_{(F)}^S}$ will not give rise to any wave equation. Using Eq.~\eqref{eq:adot} and \eqref{eq:VFs}, we find that the evolution of $\mathcal{V}\indices{_{(F)}^S}$ is simply determined by the scale factors on the background
\begin{equation}
\begin{aligned}
\frac{\dot{\mathcal{V}}\indices{_{(F)}^S}}{\mathcal{V}\indices{_{(F)}^S}} = -\left(\Sigma + \frac{\Theta}{3}\right)  = -\frac{\dot{a}_1}{a_1} \implies \mathcal{V}\indices{_{(F)}^S}= \frac{\mathcal{V}\indices{_{(F)}^S}(t_0)}{a_1}, \quad \textnormal{if}\quad a_1(t_0) = 1.
\end{aligned}
\end{equation}
Thus, to obtain any interesting wave equations, we instead look at $\mathcal{V}\indices{_{(F)}_*^V}$ and $\overline{\mathcal{V}}\indices{_{(F)}_*^V}$. In doing so, we will assume that the plasma is ideal ($\eta \rightarrow 0$), and that the wave lengths considered are much smaller than the scales on the background, so that
\begin{equation}
k^2, \frac{\pkk}{a_1^2}, \frac{\okk}{a_2^2} \gg \Sigma^2, \Theta^2, \mathfrak{B}^2, \mu_{(F)}, \mathcal{E}.
\end{equation}  
We will then assume that the frequencies for the variables $\mathcal{V}\indices{_{(F)}_*^V}$, $\overline{\mathcal{V}}\indices{_{(F)}_*^V}$, $\mathfrak{E}\indices{_*^V}$, $\overline{\mathfrak{E}}\indices{_*^V}$,  $\mathcal{H}\indices{_*^T}$ and $\overline{\mathcal{H}}\indices{_*^T}$ are of order $k$ in this limit, so that, when applying a time derivative on one of these coefficients, the result is one order higher in $k$. The idea is then to truncate the equations, keeping only the terms with the highest orders in $k$. However, to be able to compare the orders of various terms in the equations, we first have to consider the inherent difference in order between the harmonic coefficients. 

 With the chosen initial conditions in this section, $\Omega^S, \overline{\Omega}\indices{_*^V}$ and $\mathcal{V}\indices{_{(F)}^S}$ are identically zero, and will hence be omitted in the following. Motivated by numerical results using the aforementioned initial conditions, we will treat the magnitudes of $\overline{\mathcal{H}}\indices{_*^T}, \mathcal{E}\indices{_*^T}, \overline{\mathfrak{E}}\indices{_*^V}$ and $\mathcal{V}\indices{_{(F)}_*^V}$ as being of the same order in $k$, up to factors of order $10$. The coefficients $\mu\indices{_{(F)}_*^V}$ and $\mathcal{B}\indices{_*^V}$ are in turn treated as being one order higher in $k$ whilst $\Sigma\indices{_*^T}$ is seen to be one order lower in $k$ than $\mathcal{V}\indices{_{(F)}_*^V}$. With a similar numerical motivation, we assume that the magnitudes of $\overline{\mathcal{E}}\indices{_*^T}, \mathcal{H}\indices{_*^T}, \mathfrak{E}\indices{_*^V}, \overline{\mathfrak{B}}\indices{_*^V} $ and $\mathcal{V}\indices{_{(F)}_*^V}$ are all of the same order in $k$, whilst the magnitude of $\mathfrak{E}^S$ is significantly smaller. The smallness of $\mathfrak{E}^S$ is consistent with the ideal limit $\eta \rightarrow 0$, as $\mathfrak{E}^S$ should then tend to zero.  
 
 On using these relative orders in $k$ and keeping only the terms of highest order in $k$, the following wave equations can be derived
 \begin{align}
\ddot{\mathcal{V}}\indices{_{(F)}_*^V}  + {v_A}^2\left(\frac{\pkk}{a_1^2} + \frac{\okk}{a_2^2}\right)\mathcal{V}\indices{_{(F)}_*^V} &= 0, \label{eq:VFVwave2}\\
~
\ddot{\overline{\mathcal{V}}}\indices{_{(F)}_*^V} + {v_A}^2\frac{\pkk}{a_1^2}\overline{\mathcal{V}}\indices{_{(F)}_*^V} &= 0, \label{eq:VFbarVwave2}
\end{align}
where the previously defined Alfvén velocity $v_A$ appears. When deriving Eqs.~(\ref{eq:VFVwave2}) and (\ref{eq:VFbarVwave2}), we used Eqs.~(\ref{ps:eq56}) and (\ref{ps:eq23}) to write the currents in Eqs.~(\ref{eq:VFbarV}) and (\ref{eq:VFV}) in terms of derivatives of the electric field. The electric field coefficients were then eliminated by noting that, in the ideal limit $\eta\rightarrow 0$, we must have
\begin{align}
\mathfrak{E}^S &\rightarrow 0, \label{eq:idealEs} \\
~
\mathfrak{E}\indices{_*^V} &\rightarrow \mathfrak{B}\overline{\mathcal{V}}\indices{_{(F)}_*^V}, \label{eq:idealEv}\\
~
\overline{\mathfrak{E}}\indices{_*^V} &\rightarrow - \mathfrak{B}\mathcal{V}\indices{_{(F)}_*^V} \label{eq:idealEbarv}, 
\end{align}
for the currents in Eqs.~(\ref{eq:Current1})–(\ref{eq:Current2}) and (\ref{eq:Js}) to remain finite. 

The wave equations for the plasma velocities can be compared with non-relativistic results for ideal MHD waves in magnetized plasmas \cite{Baumjohann_2012}. It can then be seen that Eq.~(\ref{eq:VFVwave2}) is consistent with a fast magnetosonic mode in the limit of vanishing speed of sound. The fact that $\mathcal{V}\indices{_{(F)}^S}$ shows no wave-like behavior is also consistent with a slow magnetosonic mode in this limit. This is reasonable since we have neglected the fluid pressures, and hence also the acoustic modes. As for $\overline{\mathcal{V}}\indices{_{(F)}^V}$, Eq.~(\ref{eq:VFbarVwave2}) is seen to be consistent with a shear Alfvén mode. 

Continuing with the tensor coefficients of the magnetic part of the Weyl tensor, these can, to highest order in $k$, be found to satisfy the equations 
\begin{align}
 \ddot{\overline{\mathcal{H}}}\indices{_*^T} +k^2 \overline{\mathcal{H}}\indices{_*^T} &= \frac{i\pk\ok}{a_1 a_2}\left(\mu_{(F)} + 2\mathfrak{B}^2\right)\mathcal{V}\indices{_{(F)}_*^V}, \\
 ~ 
\ddot{\mathcal{H}}\indices{_*^T} +k^2\mathcal{H}\indices{_*^T} &=  \frac{i\pk\ok}{a_1a_2} \left({\mu_{(F)}}+ 2\mathfrak{B}^2\right) {\overline{\mathcal{V}}_{(F)}}\indices{_*^V} .\label{eq:HTwave}
 \end{align}
Assuming that ${\overline{\mathcal{V}}_{(F)}}\indices{_*^V}(t_0)$ and ${\mathcal{V}_{(F)}}\indices{_*^V}(t_0)$ are real constants,  $\mathcal{H}\indices{_*^T}(t_0) = \overline{\mathcal{H}}\indices{_*^T}(t_0) = 0$, and that the scale factors and the background variables are approximately constant on the time scales that we are considering,  Eqs.~(\ref{eq:VFVwave2})–(\ref{eq:HTwave}) have the solutions
\begin{align}
{\mathcal{V}_{(F)}}\indices{_*^V} &= {\mathcal{V}_{(F)}}\indices{_*^V}(t_0)\cos\left(v_Akt\right) , \quad {\overline{\mathcal{V}}_{(F)}}\indices{_*^V} = {\overline{\mathcal{V}}_{(F)}}\indices{_*^V}(t_0)\cos\left(v_Ak_pt\right), \\
~
\overline{\mathcal{H}}\indices{_*^T} &= \frac{2i\pk\ok\left(\mu_{(F)}+2\mathfrak{B}^2 \right){\mathcal{V}_{(F)}}\indices{_*^V}(t_0)}{k^2a_1a_2\left(1-v_A^2\right)} \sin\left(\frac{1-v_A}{2}kt\right)\sin\left(\frac{1+v_A}{2}kt\right), \label{eq:barHTwave} \\
~
 \mathcal{H}\indices{_*^T} &= \frac{2i\pk\ok\left(\mu_{(F)}+2\mathfrak{B}^2 \right){\overline{\mathcal{V}}_{(F)}}\indices{_*^V}(t_0)}{k^2a_1a_2\left(1-\left(v_Ak_p/k\right)^2\right)} \sin\left(\frac{1-\left(v_Ak_p/k\right)}{2}kt\right)\sin\left(\frac{1+\left(v_Ak_p/k\right)}{2}kt\right).
\end{align}
Here $k_p = \lvert\pk/a_1\rvert$, and all scale factors and background quantities should be evaluated at $t_0 = 0$. It should here be noted that we have set the first order derivatives to zero at $t_0$. This is consistent with the first order differential equations, as they imply that the derivatives of these harmonic coefficients are of order $k^0$ at $t_0$. As the amplitudes of the derivatives should be of order $k$, their initial values of order $k^0$ are neglected to highest order. 

Studying the solutions for $ \mathcal{H}\indices{_*^T}$ and $\overline{\mathcal{H}}\indices{_*^T}$ we clearly see the cause of the beat pattern in Fig.~\ref{fig:3},  which emerges due to the interaction between gravitational wave modes and the magnetized MHD waves. It should also be noted that a resonance can occur when the Alfvén velocity approaches unity. For $\overline{\mathcal{H}}\indices{_*^T}$, the resonance is independent of the precise relation between $\pk$ and $\ok$, whereas the factor $k_p/k$ limits the resonance for  $\mathcal{H}\indices{_*^T}$.  

\begin{figure}[h]
\centering
\includegraphics[scale=1]{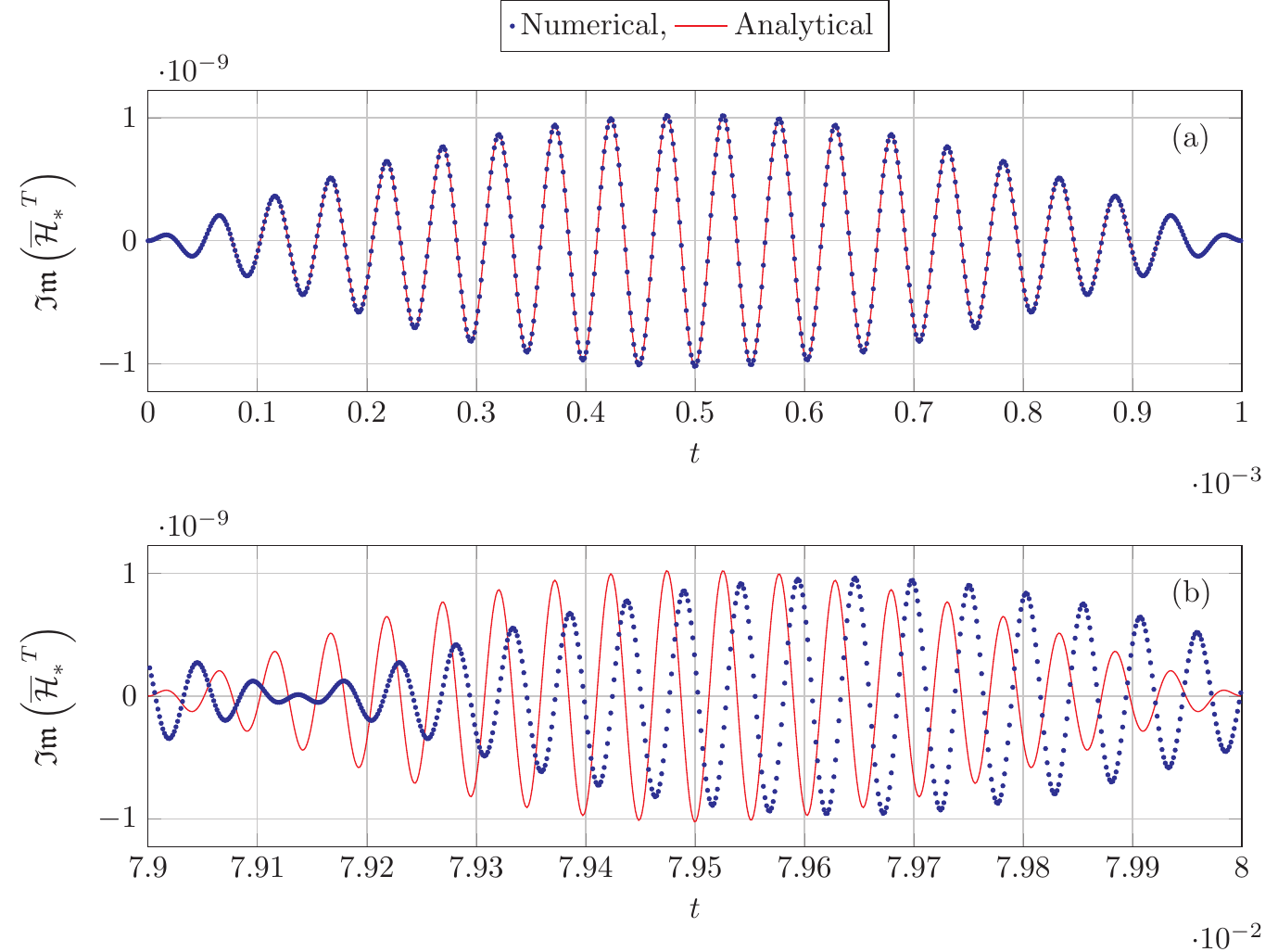}
\caption{Comparison between the analytical expression and the numerical results for the magnetic part of the Weyl tensor in the even sector. The perturbation length scale was chosen as $L = 10^{-5}L_H$, and the background was specified using Eq.~(\ref{eq:InitialBackground}). The initial magnitude of the magnetic field on the background was chosen so that $v_A(t_0) = 0.95$. As for the initial perturbations, these were chosen according to Eqs.~(\ref{eq:InitialPert4start})–(\ref{eq:InitialPert4end}). In all calculations, $\pk = 2\ok = 2k(t_0)/\sqrt{5}$ and $\eta(t_0) = 10^{-12}$.} 
\label{fig:4}
\end{figure}

Comparing the analytical solution for $\overline{\mathcal{H}}\indices{_*^T}$ with the numerical data, we see a good agreement initially in Fig.~\ref{fig:4}a. However, in Fig.~\ref{fig:4}b it can be seen that, as time increases, the numerical results become shifted relative to the analytical expression, indicating a change of frequency for the numerical results. The amplitude of the numerical results also seems to decrease slightly in comparison to the analytical expression, but this effect is less noticeable than the frequency shift for the chosen parameters. These differences between the numerical and analytical results are expected, as we, when deriving the analytical expression, have neglected frequency redshifts and damping effects due to the evolution of the background spacetime.   
     
\section{Discussion}

Some of the numerical results presented here agree rather well with analytical expressions presented in \cite{Zunckel_2006}, which were found using second order gauge invariant perturbation theory on spatially flat FLRW backgrounds in the ideal and cold MHD limit. In accordance with Eq.~(50) in \cite{Zunckel_2006} it was there found that, in the superhorizon limit, the dominant late-time contribution to the generated magnetic field perturbation due to gravitational waves depended quadratically on the ratio between the length scale of the magnetized region to first order and the Hubble scale. This contribution was also found to be proportional to the initial shear perturbation \cite{Zunckel_2006} . Similar quadratic dependencies on the characteristic length scale of the magnetic field have also been reported in \cite{Tsagas_2003, Betschart_2005}.  Replacing the length scale of the magnetized region in \cite{Zunckel_2006} with the perturbation length scale $L$ used here, we find, despite the differences in approach and background spacetimes, a similar quadratic dependence in Figs.~\ref{fig:1}a–c, where we have assumed that the initial tensorial shear perturbations are independent of the length scale. It should also be noted that the final result in \cite{Zunckel_2006} for the generated magnetic field perturbation, which was obtained by integrating the equations, has been argued to not be gauge invariant \cite{Mongwane_2012}.  Since our generated gauge invariant magnetic field variables $\overline{\mathfrak{B}}\indices{_*^V}, \mathfrak{B}\indices{_*^V}$, and $a_2k_{\perp}^{-1}\mathcal{B}\indices{_*^V}$ show similar quadratic dependencies on the characteristic length scale, despite the differences, this result seems rather robust and independent of the precise details. 

However, focusing on the subhorizon scale, we note some differences between our results and Eq.~(50) in \cite{Zunckel_2006}. Setting the initial velocity perturbation to zero in Eq.~(50) from \cite{Zunckel_2006}, it  predicts that the dominant late time contribution in the subhorizon limit should depend linearly on the characteristic length scale of the magnetic field. When including velocity perturbations, the dominant contribution is instead predicted to be proportional to the initial velocity perturbation and inversely proportional to the length scale of the magnetic field \cite{Zunckel_2006}. However, with the initial conditions, integration times, and wave lengths considered here, the generated magnetic fields in Figs.\ref{fig:1}a–c and g–i are found to be rather independent of the length scale of the perturbations in the subhorizon limit. Although the values at $t=8$ oscillate in this limit, the general trend of these values and the calculated maximum amplitudes seem to be independent of $L/L_H$. When considering Figs.\ref{fig:1}d–f, it must be noted that $B(t_0)$ introduced in the initial conditions for this case contributes with a factor $L^{-2}$ through the initial velocity perturbation in the subhorizon limit. Therefore, if we assume that the generated field perturbation is proportional to the initial velocity perturbation and remove the additional factors of $L$ due to $B(t_0)$ by multiplying with $L^2$, we would expect the maximum values to be almost independent of the length scale in this limit whilst the values at $t = 8$ show a slope of unity.  These subhorizon results are in contrast to the predictions from \cite{Zunckel_2006}, from which we would expect Figs.\ref{fig:1}a–c to show a slope of unity and Figs.\ref{fig:1}g–i a slope of negative unity. However, looking at superhorizon scales in Figs.~\ref{fig:1}g–i we see the inverse dependence on length scale that \cite{Zunckel_2006} predicts for subhorizon scales. This may be due to that we, in some sense, have removed some gravitational effects from the initial conditions for Figs.~\ref{fig:1}g–i, so that the contributions from these effects no longer overshadow the velocity contribution in the superhorizon limit. 

The exact explanation behind the difference between our subhorizon results and those in \cite{Zunckel_2006} is however hard to pinpoint without detailed analytical solutions. Possible, but speculative, explanations could be differences in the studied backgrounds, the initial conditions for the perturbations, the chosen model for the electrical resistivity, or that the wave lengths studied here were simply too large. It may also be due to some decaying non-dominant terms, similar to the implicit terms in \cite{Zunckel_2006}, that have not yet completely vanished at the integration times that our results were extracted. Due to all the possible differences, it is not surprising that our results differ from those in \cite{Zunckel_2006}. It is, however, interesting that they qualitatively still agree rather well in the superhorizon limit. 

Finally, it should be noted that the interactions we observe are crucially dependent on the background magnetic field. As an example, it can be seen in Eq.~(\ref{eq:barHTwave}) that the magnitude of the generated perturbations in the magnetic part of the Weyl tensor increases with increasing $\mathfrak{B}(t_0)$, especially as $v_A \rightarrow 1$. Furthermore, looking at the equations in Sec.~\ref{sec:Final System}, we note that if we were to set $\mathfrak{B} = 0$, this would imply that the equations for the electromagnetic fields decouple from the others, at least to first order. Thus, the interactions we see here to first order are fundamentally dependent on the anisotropic nature of the background spacetime, as this is what allows us to have a non-zero background magnetic field.

\section{Conclusions}
In this paper, we have studied linear electromagnetic, gravitational, and plasma related perturbations on homogeneous and orthogonal LRS class II spacetimes, where we have allowed for a non-zero magnetic field on the background. Using a $1+1+2$ covariant formalism and ultimately applying the cold MHD limit, we have derived a closed set of equations governing the harmonic coefficients of the perturbations. On analyzing the system, we have observed the generation of magnetic field perturbations due to initial perturbations in tensorial quantities related to gravitational waves. The generated fields were then found to depend quadratically on the characteristic length scale in the superhorizon limit, in agreement with previous works using FLRW backgrounds. However, in the subhorizon limit, the similarities were not as apparent. 

In addition to the generation of magnetic field perturbations, we also observed beat waves arising due to interference between gravitational waves and the magnetized MHD modes. Thus, replacing the background FLRW spacetime with an anisotropic LRS class II spacetime and allowing for a zeroth order magnetic field,  we were indeed able to see interesting interactions already to first order in the perturbations. 

\begin{appendices}
\renewcommand{\theequation}{A.\arabic{equation}}
\setcounter{equation}{0}
\section{Commutation Relations}\label{sec:Commutation Relations}
Here we state the commutation relations between the projected derivatives to first order. These relations are crucial when writing the first order equations in terms of the gauge invariant variables. For a general scalar $\Psi$ that is non-zero to zeroth order, we have
\begin{align}
\hat{\dot{\Psi}} - \dot{\hat{\Psi}} =& -\mathcal{A}\dot{\Psi} + \left(\Sigma + \frac{\Theta}{3}\right)\hat{\Psi},\\
~
\tensor{\delta}{_a}\dot{\Psi} - \tensor{N}{_a^b}\big(\tensor{\delta}{_b}\Psi\dot{\big)} =& -\tensor{\mathcal{A}}{_a}\dot{\Psi} -\frac{1}{2}\left(\Sigma - \frac{2\Theta}{3}\right)\tensor{\delta}{_a}\Psi, \\
~
\tensor{\delta}{_a}\hat{\Psi} - \tensor{N}{_a^b}\big(\tensor{\delta}{_b}\Psi\hat{\big)} = &-2\tensor{\epsilon}{_a_b}\tensor{\Omega}{^b}\dot{\Psi}, \\
~
 \tensor{\delta}{_{[a}}\tensor{\delta}{_{b]}}\Psi =&~ \Omega\tensor{\epsilon}{_a_b}\dot{\Psi}.
\end{align}
For 2-vectors $\Psi_a$ we similarly have  
\begin{align}
\tensor{\hat{\dot{\Psi}}}{_{\bar{a}}} -\tensor{\dot{\hat{\Psi}}}{_{\bar{a}}} = &~\left(\Sigma+ \frac{\Theta}{3}\right)\tensor{\hat{\Psi}}{_{\bar{a}}},\\
~
\tensor{\delta}{_a}\tensor{\dot{\Psi}}{_b} - \tensor{N}{_a^c}\tensor{N}{_b^d}\big(\tensor{\delta}{_c}\tensor{\Psi}{_d}\dot{\big)} =& -\frac{1}{2}\left(\Sigma - \frac{2\Theta}{3}\right)\tensor{\delta}{_a}\tensor{\Psi}{_b},\\
~
\tensor{\delta}{_a}\tensor{\hat{\Psi}}{_b} - \tensor{N}{_a^c}\tensor{N}{_b^d}\big(\tensor{\delta}{_c}\tensor{\Psi}{_d}\hat{\big)} =&~  0, \\
~
\tensor{\delta}{_{[a}}\tensor{\delta}{_{b]}}\tensor{\Psi}{_c} =&~ \frac{\mathcal{R}}{2}\tensor{N}{_c_{[a}}\tensor{\Psi}{_{b]}},
\end{align}
Finally, for general PSTF 2-tensors $\Psi_{ab}$, it holds that
\begin{align}
\tensor{\hat{\dot{\Psi}}}{_{\{a}_{b\}}} - \tensor{\dot{\hat{\Psi}}}{_{\{a}_{b\}}} =&~ \left(\Sigma + \frac{\Theta}{3}\right)\tensor{\hat{\Psi}}{_{\bar{a}}_{\bar{b}}}, \\
~
\tensor{\delta}{_a}\tensor{\dot{\Psi}}{_b_c} - \tensor{N}{_a^d}\tensor{N}{_b^e}\tensor{N}{_c^f}\big(\tensor{\delta}{_d}\tensor{\Psi}{_e_f}\dot{\big)} =& - \frac{1}{2}\left(\Sigma - \frac{2\Theta}{3}\right)\tensor{\delta}{_a}\tensor{\Psi}{_b_c}, \\
~
\tensor{\delta}{_a}\tensor{\hat{\Psi}}{_b_c} - \tensor{N}{_a^d}\tensor{N}{_b^e}\tensor{N}{_c^f}\big(\tensor{\delta}{_d}\tensor{\Psi}{_e_f}\hat{\big)} = &~ 0, \\
~
\tensor{\delta}{_{[a}}\tensor{\delta}{_{b]}}\tensor{\Psi}{^c^d} = &~ \mathcal{R}\tensor{N}{^{(c}_{[a}}\tensor{\Psi}{_{b]}^{d)}}.
\end{align}

\renewcommand{\theequation}{B.\arabic{equation}}
\setcounter{equation}{0}
\section{Properties of the Harmonics}\label{sec:Properties of the Harmonics}
We here state some relations for the harmonic functions that are useful when performing the harmonic decomposition of the linearized equations. The even and odd vector harmonics satisfy the relations
\begin{equation}
\tensor{N}{^a^b}\tensor{{Q^{k_{\perp}}}}{_a}\tensor{{\overline{Q}^{k_{\perp}}}}{_b} = 0, \\
\end{equation}
\begin{equation}
{\dot{Q}}^{k_{\perp}}\tensor{}{_a}  = {\hat{Q}}^{k_{\perp}}\tensor{}{_a} = {\dot{\overline{Q}}}^{k_{\perp}}\tensor{}{_a}  = {\hat{\overline{Q}}}^{k_{\perp}}\tensor{}{_a} = 0,
\end{equation}
\begin{alignat}{3}
\tensor{{Q^{k_{\perp}}}}{_a} &= -\tensor{\epsilon}{_a^b}\tensor{{\overline{Q}^{k_{\perp}}}}{_b} , \quad & \tensor{{\overline{Q}^{k_{\perp}}}}{_a} &= \tensor{\epsilon}{_a^b}\tensor{{Q^{k_{\perp}}}}{_b} , \\
~
\delta^2\tensor{{Q^{k_{\perp}}}}{_a} &= \frac{\mathcal{R}a_2^2- 2\okk}{2a_2^2}\tensor{{Q^{k_{\perp}}}}{_a}, \quad & \delta^2\tensor{{\overline{Q}^{k_{\perp}}}}{_a} &= \frac{\mathcal{R}a_2^2- 2\okk}{2a_2^2}\tensor{{\overline{Q}^{k_{\perp}}}}{_a}, \\
~
\tensor{\delta}{^a}\tensor{{Q^{k_{\perp}}}}{_a} &= -\frac{\okk}{a_2}{Q^{k_{\perp}}}, \quad & \tensor{\delta}{^a}\tensor{{\overline{Q}^{k_{\perp}}}}{_a} &= 0, \\
~
\tensor{\epsilon}{^a^b}\tensor{\delta}{_a}\tensor{{Q^{k_{\perp}}}}{_b} &= 0, \quad& \tensor{\epsilon}{^a^b}\tensor{\delta}{_a}\tensor{{\overline{Q}^{k_{\perp}}}}{_b} &= \frac{\okk}{a_2}{Q^{k_{\perp}}},
\end{alignat}

whereas for the tensor harmonics, we have
\begin{equation}
\tensor{N}{^a^b}\tensor{N}{^c^d}\tensor{{Q^{k_{\perp}}}}{_a_c}\tensor{{\overline{Q}^{k_{\perp}}}}{_b_d}= 0,
\end{equation}
\begin{equation}
{\dot{Q}}^{k_{\perp}}\tensor{}{_a_b}  = {\hat{Q}}^{k_{\perp}}\tensor{}{_a_b} = {\dot{\overline{Q}}}^{k_{\perp}}\tensor{}{_a_b}  = {\hat{\overline{Q}}}^{k_{\perp}}\tensor{}{_a_b} = 0,
\end{equation}
\begin{alignat}{2}
\tensor{{Q^{k_{\perp}}}}{_a_b} &= \tensor{\epsilon}{_{\{a}^c}\tensor{{\overline{Q}^{k_{\perp}}}}{_{b\}}_c}, \quad& \tensor{{\overline{Q}^{k_{\perp}}}}{_a_b} &= -\tensor{\epsilon}{_{\{a}^c}\tensor{{Q^{k_{\perp}}}}{_{b\}}_c}, \\
~
\delta^2\tensor{{Q^{k_{\perp}}}}{_a_b} &= \frac{2\mathcal{R}a_2^2 -\okk}{a_2^2}\tensor{{Q^{k_{\perp}}}}{_a_b}, \quad& \delta^2\tensor{{\overline{Q}^{k_{\perp}}}}{_a_b} &= \frac{2\mathcal{R}a_2^2 -\okk}{a_2^2}\tensor{{\overline{Q}^{k_{\perp}}}}{_a_b}, \\
~
\tensor{\delta}{^b}\tensor{{Q^{k_{\perp}}}}{_a_b} &= \frac{\mathcal{R}a_2^2 - \okk}{2a_2}\tensor{{Q^{k_{\perp}}}}{_a} , \quad& \tensor{\delta}{^b}\tensor{{\overline{Q}^{k_{\perp}}}}{_a_b} &=- \frac{\mathcal{R}a_2^2 - \okk}{2a_2}\tensor{{\overline{Q}^{k_{\perp}}}}{_a} , \\
~
\tensor{\epsilon}{_a^c}\tensor{\delta}{^b}\tensor{{Q^{k_{\perp}}}}{_b_c} &= \frac{\mathcal{R}a_2^2- \okk}{2a_2}\tensor{{\overline{Q}^{k_{\perp}}}}{_a}, \quad & \tensor{\epsilon}{_a^c}\tensor{\delta}{^b}\tensor{{\overline{Q}^{k_{\perp}}}}{_b_c} &= \frac{\mathcal{R}a_2^2- \okk}{2a_2}\tensor{{Q^{k_{\perp}}}}{_a}, \\
~
\tensor{\epsilon}{^b^c}\tensor{\delta}{_b}\tensor{{Q^{k_{\perp}}}}{_a_c} &= \frac{\mathcal{R}a_2^2- \okk}{2a_2}\tensor{{\overline{Q}^{k_{\perp}}}}{_a}, \quad & \tensor{\epsilon}{^b^c}\tensor{\delta}{_b}\tensor{{\overline{Q}^{k_{\perp}}}}{_a_c} &= \frac{\mathcal{R}a_2^2- \okk}{2a_2}\tensor{{Q^{k_{\perp}}}}{_a}.
\end{alignat}

\renewcommand{\theequation}{C.\arabic{equation}}
\setcounter{equation}{0}
\section{Harmonic Coefficients in the Multifluid System}\label{sec:Harmonic Coefficients in the Multifluid System}
Here we state the harmonic decompositions of the linearized equations. Since the equations originating from the Ricci and Bianchi identities can be found in \cite{Bradley_2021}, we here only show the additional equations pertaining to the decomposition of the total energy-momentum tensor, energy-momentum and particle conservation for the individual plasma components, and Maxwell's equations.

\subsection{Decomposition of the Total Energy-Momentum Tensor}
\begin{align}
\mu^V &= \sum_{i}\mu\indices{_{(i)}^V} + \mathfrak{B}\mathcal{B}^V, \\
~
\overline{\mu}^V &= \sum_{i}\overline{\mu}\indices{_{(i)}^V} + \mathfrak{B}\overline{\mathcal{B}}^V, \\
~
p^V &= \sum_{i}p\indices{_{(i)}^V} + \frac{\mathfrak{B}}{3}\mathcal{B}^V,\\
~
\overline{p}^V &= \sum_{i}\overline{p}\indices{_{(i)}^V} + \frac{\mathfrak{B}}{3}\overline{\mathcal{B}}^V, \\
~
Q^S &= \sum_{i}\left(\mu_{(i)} + p_{(i)}\right)\mathcal{V}\indices{_{(i)}^S}, \\
~
Q^V &= \sum_{i}\left(\mu_{(i)} + p_{(i)}\right)\mathcal{V}\indices{_{(i)}^V} -  \mathfrak{B}\overline{\mathfrak{E}}^V, \\
~
\overline{Q}^V &= \sum_{i}\left(\mu_{(i)} + p_{(i)}\right)\overline{\mathcal{V}}\indices{_{(i)}^V} +  \mathfrak{B}\mathfrak{E}^V, \\
~
Y^V &= -\frac{4\mathfrak{B}}{3}\mathcal{B}^V, \\
~
\overline{Y}^V &= -\frac{4\mathfrak{B}}{3}\overline{\mathcal{B}}^V, \\
~
\Pi^V &= -\mathfrak{B}\mathfrak{B}^V, \\
~
\overline{\Pi}^V &= -\mathfrak{B}\overline{\mathfrak{B}}^V.
\end{align}

\subsection{Equations for Each Fluid Component}

\begin{align}
\begin{split}
 \dot{\mu}\indices{_{(i)}^V} = &~\mathcal{A}^V\dot{\mu}_{(i)} + \frac{1}{2}\left(\Sigma - \frac{2\Theta}{3}\right)\mu\indices{_{(i)}^V} - \left(\mu\indices{_{(i)}^V} +p\indices{_{(i)}^V}\right)\Theta\\
~
& - \left(\mu_{(i)}+p_{(i)}\right)\left(\frac{i\pk}{a_1a_2}\mathcal{V}\indices{_{(i)}^S} -\frac{\okk}{a_2^2}\mathcal{V}\indices{_{(i)}^V}+  W^V\right) + \frac{1}{a_2}\varepsilon\indices{_{(i)}^S},
\end{split}\\
~
\begin{split}
 \dot{\overline{\mu}}\indices{_{(i)}^V} = &~\overline{\mathcal{A}}^V\dot{\mu}_{(i)} + \frac{1}{2}\left(\Sigma - \frac{2\Theta}{3}\right)\overline{\mu}\indices{_{(i)}^V} - \left(\overline{\mu}\indices{_{(i)}^V} +\overline{p}\indices{_{(i)}^V}\right)\Theta\\
~
& - \left(\mu_{(i)}+p_{(i)}\right)\overline{W}^V, 
\end{split}\\
~
\begin{split}
\left(\mu_{(i)} + p_{(i)}\right)\dot{\mathcal{V}}\indices{_{(i)}^S} = &-\frac{ia_2\pk}{a_1}p\indices{_{(i)}^V} -\left( 2a_2\overline{\Omega}^V+\mathcal{V}\indices{_{(i)}^S}\right)\dot{p}_{(i)}  + \rho_{(i)}\mathfrak{E}^S \\
~
&- \left(\mu_{(i)} + p_{(i)}\right)\left( \mathcal{A}^S + \left(\Sigma + \frac{\Theta}{3}\right)\mathcal{V}\indices{_{(i)}^S}\right)+ \mathcal{F}\indices{_{(i)}^S},
\end{split}\\
~
\begin{split}
0 = &-\frac{ia_2\pk}{a_1}\overline{p}\indices{_{(i)}^V} + 2a_2\dot{p}_{(i)}\Omega^V, 
\end{split}\\
~
\begin{split}
\left(\mu_{(i)} + p_{(i)}\right)\dot{\mathcal{V}}\indices{_{(i)}^V} = &-\dot{p}_{(i)}\mathcal{V}\indices{_{(i)}^V} - p\indices{_{(i)}^V} + \rho_{(i)}\left( \mathfrak{E}^V - \mathfrak{B}\overline{\mathcal{V}}\indices{_{(i)}^V}\right) \\
~
&-\left(\mu_{(i)} + p_{(i)}\right)\left( \mathcal{A}^V - \frac{1}{2}\left(\Sigma-\frac{2\Theta}{3} \right)\mathcal{V}\indices{_{(i)}^V} \right) + \mathcal{F}\indices{_{(i)}^V},
\end{split}\\
~
\begin{split}
\left(\mu_{(i)} + p_{(i)}\right)\dot{\overline{\mathcal{V}} }\indices{_{(i)}^V} = &-\dot{p}_{(i)}\overline{\mathcal{V}}\indices{_{(i)}^V} - \overline{p}\indices{_{(i)}^V} + \rho_{(i)}\left( \overline{\mathfrak{E}}^V + \mathfrak{B}\mathcal{V}\indices{_{(i)}^V} \right) \\
~
&-\left(\mu_{(i)} + p_{(i)}\right)\left( \overline{\mathcal{A}}^V - \frac{1}{2}\left(\Sigma-\frac{2\Theta}{3} \right)\overline{\mathcal{V}}\indices{_{(i)}^V} \right) + \overline{\mathcal{F}}\indices{_{(i)}^V},
\end{split}\\
~
\begin{split}
\dot{\mathcal{Z}}\indices{_{(i)}^V} = &~\mathcal{A}^V\dot{\mathcal{N}}_{(i)} + \left(\frac{\Sigma}{2} - \frac{4\Theta}{3}\right)\mathcal{Z}\indices{_{(i)}^V}  \\
~
&- \mathcal{N}_{(i)}\left(W^V + \frac{i\pk}{a_1a_2}\mathcal{V}\indices{_{(i)}^S} -\frac{\okk}{a_2^2}\mathcal{V}\indices{_{(i)}^V}\right),
\end{split}\\
~
\begin{split}
\dot{\overline{\mathcal{Z}} } \indices{_{(i)}^V} = &~\overline{\mathcal{A}}^V\dot{\mathcal{N}}_{(i)} + \left(\frac{\Sigma}{2} - \frac{4\Theta}{3}\right)\overline{\mathcal{Z}} \indices{_{(i)}^V} - \mathcal{N}_{(i)}\overline{W}^V.
\end{split}
\end{align}

\subsection{Maxwell's Equations}

\begin{align}
\dot{\mathfrak{E}}^S - \frac{\okk}{a_2}\overline{\mathfrak{B}}^V &= 2\mathfrak{B}\xi^S + \left(\Sigma - \frac{2\Theta}{3}\right)\mathfrak{E}^S -\mathcal{J}^S, \\
~
\frac{i\pk}{a_1}\mathfrak{E}^S -\frac{\okk}{a_2}\mathfrak{E}^V &= 2\mathfrak{B}\Omega^S + \rho^S, \\
~
\dot{\mathcal{B}}^V + \frac{\okk}{a_2^2}\overline{\mathfrak{E}}^V &= \mathcal{A}^V\dot{\mathfrak{B}} + \frac{3}{2}\left(\Sigma - \frac{2\Theta}{3}\right)\mathcal{B}^V  + \left(V^V - \frac{2}{3}W^V\right)\mathfrak{B},\\
~
\dot{\overline{\mathcal{B}}}^V &= \overline{\mathcal{A}}^V\dot{\mathfrak{B}} + \frac{3}{2}\left(\Sigma - \frac{2\Theta}{3}\right)\overline{\mathcal{B}}^V  + \left(\overline{V}^V - \frac{2}{3}\overline{W}^V\right)\mathfrak{B}, \\
~
\frac{i\pk}{a_1}\mathcal{B}^V  -\frac{\okk}{a_2^2}\mathfrak{B}^V &= -2\overline{\Omega}^V\dot{\mathfrak{B}} - \frac{1}{a_2}\mathfrak{B}\phi^S,\\
~
\frac{i\pk}{a_1}\overline{\mathcal{B}}^V &= 2\Omega^V\dot{\mathfrak{B}},\\
~
\dot{\mathfrak{E}}^V - \left(\frac{i\pk}{a_1}\overline{\mathfrak{B}}^V - \overline{\mathcal{B}}^V \right) &=  -\frac{1}{2}\left(\Sigma + \frac{4\Theta}{3}\right)\mathfrak{E}^V - \mathfrak{B}\left(\overline{\mathcal{A}}^V - \overline{a}^V\right)  - {\mathcal{J}}^V, \\
~
\dot{\overline{\mathfrak{E}}}^V + \left(\frac{i\pk}{a_1}\mathfrak{B}^V - \mathcal{B}^V \right) &=  -\frac{1}{2}\left(\Sigma + \frac{4\Theta}{3}\right)\overline{\mathfrak{E}}^V + \mathfrak{B}\left(\mathcal{A}^V - a^V\right) - \overline{\mathcal{J}}^V \\
~ 
\dot{\mathfrak{B}}^V + \frac{i\pk}{a_1}\overline{\mathfrak{E}}^V &= -\frac{1}{2}\left(\Sigma + \frac{4\Theta}{3}\right)\mathfrak{B}^V  + \mathfrak{B}\left(-\alpha^V +\Sigma^V - \overline{\Omega}^V\right), \\
~
\dot{\overline{\mathfrak{B}}}^V - \left(\frac{i\pk}{a_1}\mathfrak{E}^V - \frac{1}{a_2}\mathfrak{E}^S\right) &= -\frac{1}{2}\left(\Sigma + \frac{4\Theta}{3}\right)\overline{\mathfrak{B}}^V  + \mathfrak{B}\left(-\overline{\alpha}^V +\overline{\Sigma}^V + \Omega^V\right), \\
~
\mathcal{J}^S &= \sum_{i} q_{c(i)}\mathcal{N}_{(i)}\mathcal{V}\indices{_{(i)}^S}\\
~
{\mathcal{J}}^V &= \sum_{i} q_{c(i)}\mathcal{N}_{(i)}\mathcal{V}\indices{_{(i)}^V},\\
~
{\overline{\mathcal{J}}}^V &= \sum_{i} q_{c(i)}\mathcal{N}_{(i)}\overline{\mathcal{V}}\indices{_{(i)}^V},\\
~
\rho^S &= a_2\sum_{i} q_{c(i)}\mathcal{Z}\indices{_{(i)}^V},\\
~
0 &= \sum_{i} q_{c(i)}\overline{\mathcal{Z}} \indices{_{(i)}^V}.
\end{align}

\subsection{Odd Parts of the Gradient Variables}
Using the commutation relation
\begin{equation}
\tensor{\delta}{_a}\hat{\Psi} - \tensor{N}{_a^b}\big(\tensor{\delta}{_b}\Psi\hat{\big)} = -2\tensor{\epsilon}{_a_b}\tensor{\Omega}{^b}\dot{\Psi},
\end{equation}
for a scalar $\Psi$ that is non-zero to zeroth order, it should also be noted that it follows directly that 
\begin{equation}
\frac{i\pk}{a_1}\overline{\Psi}^V = 2\Omega^V\dot{\Psi}.
\end{equation} 

\renewcommand{\theequation}{D.\arabic{equation}}
\setcounter{equation}{0}
\section{Harmonic Coefficients in the Cold MHD System}\label{sec:Harmonic Coefficients in the Cold MHD System}
Here we state the equations for the harmonic coefficients after applying the cold MHD approximation. Only equations that differ from the multifluid case are presented.

The differential equations pertaining to the collective energy densities and velocities are
\begin{align}
\dot{\mu}_{(F)} = &- \mu_{(F)}\Theta, \label{eq:muF_dot}\\
~ 
 \mu =&~ \mu_{(F)}+ \frac{\mathfrak{B}^2 }{2}, \quad p =  \frac{\mathfrak{B}^2}{6}, \\
~
\begin{split}
 \dot{\mu}\indices{_{(F)}^V} = &~\mathcal{A}^V\dot{\mu}_{(F)} + \left(\frac{\Sigma}{2} - \frac{4\Theta}{3}\right)\mu\indices{_{(F)}^V} \\
~
& - \mu_{(F)}\left(\frac{i\pk}{a_1a_2}\mathcal{V}\indices{_{(F)}^S} -\frac{\okk}{a_2^2}\mathcal{V}\indices{_{(F)}^V}+  W^V\right),
\end{split} \\
~
 \dot{\overline{\mu}}\indices{_{(F)}^V} = &~\overline{\mathcal{A}}^V\dot{\mu}_{(F)} + \left(\frac{\Sigma}{2} - \frac{4\Theta}{3}\right)\overline{\mu}\indices{_{(F)}^V}  - \mu_{(F)}\overline{W}^V, \\
 ~
\mu_{(F)}\dot{\mathcal{V}}\indices{_{(F)}^S} = & - \mu_{(F)} \left( \mathcal{A}^S + \left(\Sigma + \frac{\Theta}{3}\right)\mathcal{V}\indices{_{(F)}^S}\right), \\
~ 
\mu_{(F)} \dot{\mathcal{V}}\indices{_{(F)}^V} =&  -\frac{\mathfrak{B}}{\eta}\left(\overline{\mathfrak{E}}^V + \mathfrak{B}\mathcal{V}\indices{_{(F)}^V}\right)-\mu_{(F)}\left(\mathcal{A}^V - \frac{1}{2}\left(\Sigma-\frac{2\Theta}{3} \right)\mathcal{V}\indices{_{(F)}^V}\right), \\
~ 
\mu_{(F)}\dot{\overline{\mathcal{V}}}\indices{_{(F)}^V} =&~ \frac{\mathfrak{B}}{\eta}\left(\mathfrak{E}^V - \mathfrak{B}\overline{\mathcal{V}}\indices{_{(F)}^V}\right)-\mu_{(F)}\left(\overline{\mathcal{A}}^V - \frac{1}{2}\left(\Sigma-\frac{2\Theta}{3} \right)\overline{\mathcal{V}}\indices{_{(F)}^V}\right),
\end{align}
where the currents in the momentum equations have been written as
\begin{align}
\mathcal{J}^S &= \frac{1}{\eta}\mathfrak{E}^S \label{eq:Js} \\
~
\mathcal{J}^V &= \frac{1}{\eta}\left(\mathfrak{E}^V - \mathfrak{B}\overline{\mathcal{V}}\indices{_{(F)}^V}\right),  \\
~
\overline{\mathcal{J}}^V  &= \frac{1}{\eta}\left(\overline{\mathfrak{E}}^V + \mathfrak{B}\mathcal{V}\indices{_{(F)}^V}\right), 
\end{align}
using Ohm's law.  As for the harmonic coefficients related to the decomposition of the total energy-momentum tensor, these can in turn be written as
\begin{align}
Q^S =&~ \mu_{(F)}\mathcal{V}\indices{_{(F)}^S}, \\
Q^V =&~ \mu_{(F)}\mathcal{V}\indices{_{(F)}^V} -  \mathfrak{B}\overline{\mathfrak{E}}^V, \quad \overline{Q}^V =~ \mu_{(F)}\overline{\mathcal{V}}\indices{_{(F)}^V}  +  \mathfrak{B}\mathfrak{E}^V\\
 ~
 \overline{\mu}^V =&~ \overline{\mu}\indices{_{(F)}^V} + \mathfrak{B}\overline{\mathcal{B}}^V, \quad \mu^V =~ \mu\indices{_{(F)}^V} + \mathfrak{B}\mathcal{B}^V, \\
p^V =&~\frac{\mathfrak{B}}{3}\mathcal{B}^V, \quad \overline{p}^V =~  \frac{\mathfrak{B}}{3}\overline{\mathcal{B}}^V.
\end{align}
It should be noted that the Maxwell equation involving the divergence of the electric field as well as the equations for the particle density gradients are no longer needed to close the system. The former can instead be used to determine the charge density from some specified vorticity and electric fields.  

\end{appendices}

\bibliography{ps-bibliography}

\end{document}